\documentclass[prc,twocolumn,superscriptaddress,showpacs]{revtex4}
\usepackage{graphicx,amsmath,amssymb,bm,multirow}
\newcommand{\beq}{\begin{equation}}
\newcommand{\eeq}{\end{equation}}
\newcommand{\beqa}{\begin{eqnarray}}
\newcommand{\eeqa}{\end{eqnarray}}

\begin{document}
\title{Search for three-nucleon force effects\\ 
on the longitudinal response function of $^4$He}
\author{Sonia Bacca$^{a}$\footnote{electronic address: bacca@triumf.ca}, 
Nir Barnea$^{b}$\footnote{electronic address: nir@phys.huji.ac.il},  Winfried Leidemann$^{c}$
\footnote{electronic address: leideman@science.unitn.it} and Giuseppina Orlandini$^{c}$\footnote{electronic address: orlandin@science.unitn.it}}

\affiliation{
$^{a}$TRIUMF, 4004 Wesbrook Mall,
Vancouver, B.C. V6J 2A3, Canada\\
$^{b}$Racah Institute of Physics, Hebrew University, 91904, Jerusalem,
Israel\\ 
$^{c}$Dipartimento di Fisica, Universit\`{a} di Trento and INFN\\
(Gruppo Collegato di Trento), via Sommarive 14, I-38100 Trento, Italy}

\date{\today}

\begin{abstract}
A detailed study of the $^4$He longitudinal response function $R_L(\omega,q)$ is performed 
at different kinematics, with particular emphasis on the role of three-nucleon forces. 
The effects shown are the results of  an  ab initio calculation where the full 
four-body continuum dynamics is considered via the Lorentz integral transform   method.
The contributions of the various multipoles to the longitudinal response function are analyzed and 
integral properties of the response are discussed in addition.  The Argonne V18 nucleon-nucleon interaction and 
two different three-nucleon force models (Urbana IX, Tucson-Melbourne$'$) are used. 
At lower momentum transfer ($q\leq 200$ MeV/c) three-nucleon forces play an important role.
One even finds a dependence of $R_L$ on the three-nucleon force model itself with differences up
to 10\%. Thus a Rosenbluth separation of the inclusive electron scattering cross section 
of $^4$He  at low momentum transfer would be of high value in view of a discrimination between different
three-nucleon force models. 
 
\end{abstract}

\maketitle

\section{Introduction}

An aspect of nuclear dynamics that has attracted a lot of interest in the last years is 
the importance of multi-nucleon forces and in particular of the three-nucleon force (3NF).
The nuclear potential has clearly an {\it effective} nature, therefore it is in principle 
a many-body operator. Yet the debate has concentrated for several decades mainly on its two-nucleon part.
Such a debate has taken place among the advocates of three different approaches based on
meson theory, pure phenomenology and more recently  effective field theory. Realistic potentials 
have been obtained within the three  different frameworks, relying on fits to thousands 
of N-N scattering data. As is well known such realistic potentials do not explain 
the triton binding energy and thus 3NF are necessary. Today, due to effective field theory approaches 
a new debate is taking place regarding 3NF.
However, for the determination of a  {\it realistic}  three-body potential or 
to discriminate among different models one needs to find  A$\geq3$ observables that are 3NF sensitive. 
An important activity in this direction has taken place in the last years, 
with accurate calculations of bound-state properties of nuclei of increasing mass number A~\cite{Pieper02,Navratil07}. 

We follow a complementary approach and direct our attention, instead, towards electromagnetic reactions 
in the continuum.
In fact many years of electron scattering experiments have demonstrated the power of 
electro-nuclear reactions, and in particular of the inelastic ones,  
in providing important information  on  nuclear dynamics. The possibility to vary energy $\omega$ and   
momentum $q$ transferred by the electron to the nucleus allows one to focus on different dynamical 
aspects. In fact one might find regions where the searched three-nucleon effects are sizable. 
The choice of the $^4$He target is particularly appropriate because of the following considerations.  
i) The ratio between the number of triplets and of pairs goes like (A-2)/3, therefore doubles from 
$^3$He to $^4$He.
ii) Theoretical results on hadronic scattering observables involving four 
nucleons have already shown that three-body 
effects are rather large~\cite{Kievsky08,Deltuva08}.  
iii) Since $^4$He has quite a large  average density and its binding energy per particle is similar 
to that of heavier systems, it can serve as a guideline to investigate heavier nuclei.
iv) Various {\it inclusive} $^4$He $(e,e')$ experiments have already been performed in the 
past~\cite{Carlson02}, where Rosenbluth separations have been carried out. Due to the low atomic number it is possible 
to study the longitudinal and the transverse responses separately, without the ambiguities created by the Coulomb 
distortions affecting heavier systems.
v) The Lorentz Integral Transform (LIT) method~\cite{EFROS94,REPORT07} allows to extend investigations beyond 
the three- and four-body break up thresholds. 

In this work we  concentrate on the longitudinal response function $R_L(\omega,q)$ at constant momentum transfers $q\le$ 500 MeV/c. 
Since the longitudinal response $R_L$ is much less sensitive to meson exchange
 effects  than the transverse response  
$R_T$  the use of a simple one-body 
density operator allows to concentrate on the nuclear dynamics generated by the potential.
In fact, for low $q$ two-body operators in $R_L$ are only of fourth order
in effective field theory counting (N$^3$LO) \cite{PARK03}, and their
contribution is negligible up to $q \approx$ 300 MeV/c, see Sec.~\ref{INT}.

Besides presenting new results this work gives a  more detailed analysis of those published in a previous 
Letter~\cite{PRL09}. The paper is organized as follows. In Sec.~\ref{TF} we give the definition of $R_L$ and 
explain the theoretical framework that allows to calculate it. 
In Sec.~\ref{RES} we show the results for different kinematics  and compare our results with existing data. 
In Sec.~\ref{MA} we analyze our results  as obtained  
from a multiple decomposition of the response function. 
In Sec~\ref{INT} we discuss integral properties of the longitudinal response and compare them with 
some of the results  in the literature. Finally  conclusions are drawn in Sec.~\ref{CONC}.

\section{Theoretical Framework}\label{TF}

In the one-photon-exchange approximation, the inclusive cross section
for electron scattering off a nucleus is given in terms of two
response functions, i.e.  
\begin{equation}
\frac{d^2 \sigma}{d\Omega d{\omega}}=\sigma_M\left[\frac{Q^4}{q^4}
{R_L(\omega,q)}+\left(\frac{Q^2}{2 q^2}+\tan^2{\frac{\theta}{2}}\right)
{R_T(\omega, q )}\right]
\end{equation}
where $\sigma_M$ denotes the Mott cross section, $Q^2=-q_{\mu}^2={
q}^2-\omega^2$ the squared four momentum transfer with $\omega$ and
${\bf q}$ as energy and three-momentum transfers, respectively, and
$\theta$ the electron scattering angle. The longitudinal 
and transverse response functions,
$R_L(\omega,{q})$ and $R_T(\omega,{ q})$, are determined
by the transition matrix elements of the Fourier transforms  of the 
charge and the transverse current density operators. In this work we focus on the longitudinal response
which  is given by

\beq
\label{frisp}
R_L(\omega,q)=\int \!\!\!\!\!\!\!\sum _{f} 
|\left\langle \Psi_{f}| \hat\rho({\bf q})| 
\Psi _{0}\right\rangle|
^{2}\delta\left(E_{f}+\frac{q^2}{2M}-E_{0}-\omega \right)\,, 
\eeq
where  
$M$ is the target mass, $| \Psi_{0/f} \rangle$ and 
$E_{0/f}$ denote initial and final state wave functions and energies,
respectively. The charge density operator $\rho$ is defined as
\beq
{\hat\rho}({\bf q})= \frac{e}{2} \sum_k \,(1 + \tau_k^3) \exp{[i {\bf q} \cdot {\bf r}_k]} \,, \label{rho}
\eeq
where $e$ is the proton charge and $\tau_k^3$ the isospin third component of nucleon $k$.
The $\delta$-function 
ensures energy conservation. 

As it will be clear in Sec.~\ref{RES} it is useful to consider the charge  density operator as 
decomposed  into isoscalar (S) and isovector (V) contributions 
\begin{eqnarray}
{\hat\rho}({\bf q})&=&\frac{e}{2} \sum_k \, \exp{[i {\bf q} \cdot {\bf r}_k]} + \frac{e}{2}\sum_k 
\tau_k^3 \exp{[i {\bf q} \cdot {\bf r}_k]}\nonumber\\
& \equiv & {\hat\rho}_S({\bf q})+{\hat\rho}_V({\bf q})\,.
\end{eqnarray}
Each of them  can be further decomposed into
Coulomb multipoles~\cite{EISGREI70} 
\beq
\hat{\rho}_X({\bf q})=4\pi\sum_{J\mu}
\hat{C}^{J,X}_{\mu}(q)\,Y^{J}_{\mu}(\hat{q})^*\,,\label{mult_charge}
\eeq
where the Coulomb  multipole operators $\hat{C}^{J,X}_{\mu}(q)$ are defined,
 by 
\beq
\hat{C}^{J,X}_{\mu}(q)\equiv\frac{1}{4\pi}\int d\hat{q}' \hat{\rho}_X 
({\bf q}') Y^{J}_{\mu}(\hat{q}') \,, \label{coulomb}
\eeq
with $X=S,V$ and $Y^{J}_{\mu}(\hat{q})$ denoting the spherical harmonics. 

From Eq.~(\ref{frisp}) it is evident that in principle one needs the knowledge of 
all  possible final states excited by the electromagnetic
probe, including of course states in the continuum. Thus, in a straightforward evaluation one 
would have to calculate both 
bound and continuum states. The latter constitute the major
obstacle for a many-body system, since the full  many-body scattering wave functions are not 
yet accessible for A $> $ 3. In the LIT method~\cite{EFROS94,REPORT07} this difficulty is
circumvented  by considering instead of 
$R_L(\omega,q)$ an integral transform 
${\cal L}_L(\sigma,q)$ with a Lorentzian kernel defined for a complex
parameter $\sigma=\sigma_R+i\,\sigma_I$ by
\begin{equation}
{\cal L}_L(\sigma,q)=\int d\omega
\frac{R_L(\omega,q)}
{(\omega-\sigma_R)^{2}+\sigma_I^{2}}
= \langle\widetilde{\Psi}_{\sigma,q}^{\rho}
|\widetilde{\Psi}_{\sigma,q}^{\rho}\rangle \,.\label{lorentz_transform} 
\end{equation}
The parameter $\sigma_I$ determines the resolution of the transform and is kept at a constant 
finite value ($\sigma_I\ne 0$). The basic idea of considering ${\cal L}_L$
lies in the fact that it can be evaluated from the norm of a
function $\widetilde{\Psi}_{\sigma,q}^{\rho}$,
which is the unique solution of the inhomogeneous  equation 
\begin{equation}
(\hat{H}-E_{0}-\sigma)|\widetilde{\Psi}_{\sigma,q}^{\rho}
\rangle=\hat\rho(q)|{\Psi_{0}}\rangle\,.\label{liteq}
\end{equation}
Here $\hat{H}$ denotes the nuclear Hamiltonian.
The existence of the integral in
Eq.~(\ref{lorentz_transform}) implies that  $\widetilde{\Psi}_{\sigma,q}^{\rho}$ has  
asymptotic boundary conditions similar to a bound
state. Thus, one can apply bound-state techniques for its
solution. Here we use the effective interaction hyperspherical harmonics (EIHH) method
\cite{EIHH,EIHH_3NF}.

The response function $R_L(\omega,q=const)$ is then obtained by inverting
the integral transform (\ref{lorentz_transform}). For the inversion of the LIT various
methods have been devised~\cite{EfL99,AnL05}. In particular the issue of the inversion of the 
LIT is discussed extensively in Ref.~\cite{ARXIV}.

Finally we should mention that the expression of the charge density in Eq.~(\ref{rho})
describes point particles. In order to compare our results with experimental data, after 
inversion the isoscalar 
and isovector parts of $R_L$ have to be  multiplied by the proper nucleon form factors 
\begin{equation}
\frac{1}{2}(1+\tau^3_k)\rightarrow
G_{E}^S(Q^2)+\tau^3_kG_{E}^V(Q^2)\,, 
\end{equation}
where $G_{E}^S$ and $G_{E}^V$ are the isoscalar and isovector form factors
\begin{eqnarray}
G_{E}^S=\frac{1}{2}(G_{E}^p + G_{E}^n)\,,\\
G_{E}^V=\frac{1}{2}(G_{E}^p - G_{E}^n)\,. 
\end{eqnarray}
For on-shell particles, these form factors depend on the squared four
momentum transfer $Q^2$ alone.
In principle, this is no longer 
true for the off-shell situation. However, in view of the fact that little
is known about the off-shell continuation and, furthermore, for the
moderate energy and momentum transfers considered in this work, the neglect
of such effects is justified. 
Therefore the results shown in Sec.~\ref{RES} all include the proton electric form factor with the usual
dipole parameterization 
\begin{equation}
G_{E}^p(Q^2)=G_{D}(Q^2)=\frac{1}{\left(1+\frac{Q^2}{\Lambda}\right)^2}\,
\end{equation}
($\Lambda=18.43~ {\rm fm}^{-2}$).
For the neutron electric form factor we use the parameterization from \cite{GaK71} 
\begin{equation}
G_{E}^n(Q^2)=-\frac{\mu_n \frac{Q^2}{4m^2}}{1+5.6\frac{Q^2}{4m^2}} G_{E}^p(Q^2)\,,
\end{equation}
with $\mu_n=-1.911829~\mu_N$ and $m$ being the nucleon mass.

\section{Results of the LIT calculation}\label{RES}

\begin{figure}
\includegraphics[scale=0.25,clip=]{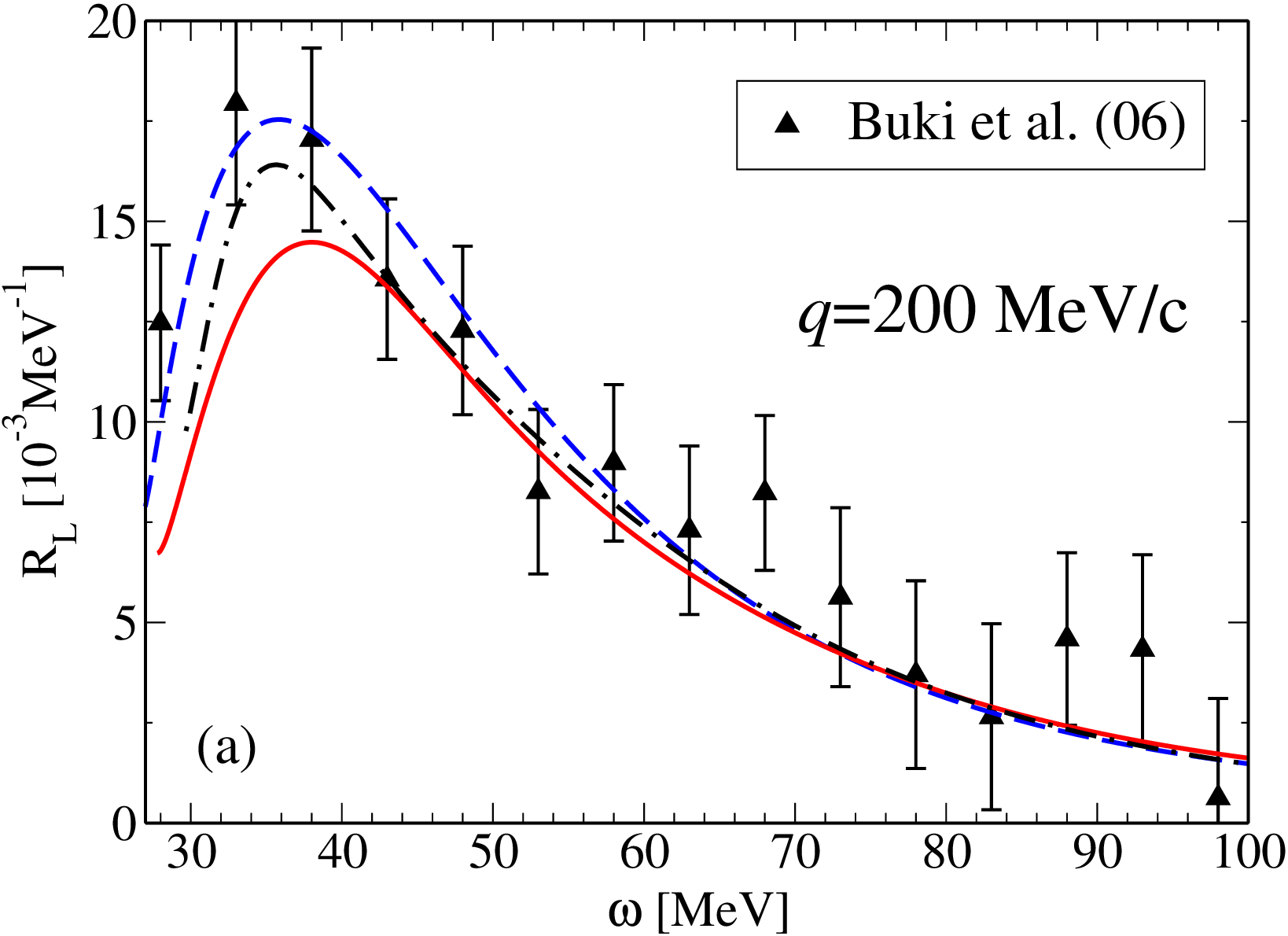}
\\
~\\
\includegraphics[scale=0.25,clip=]{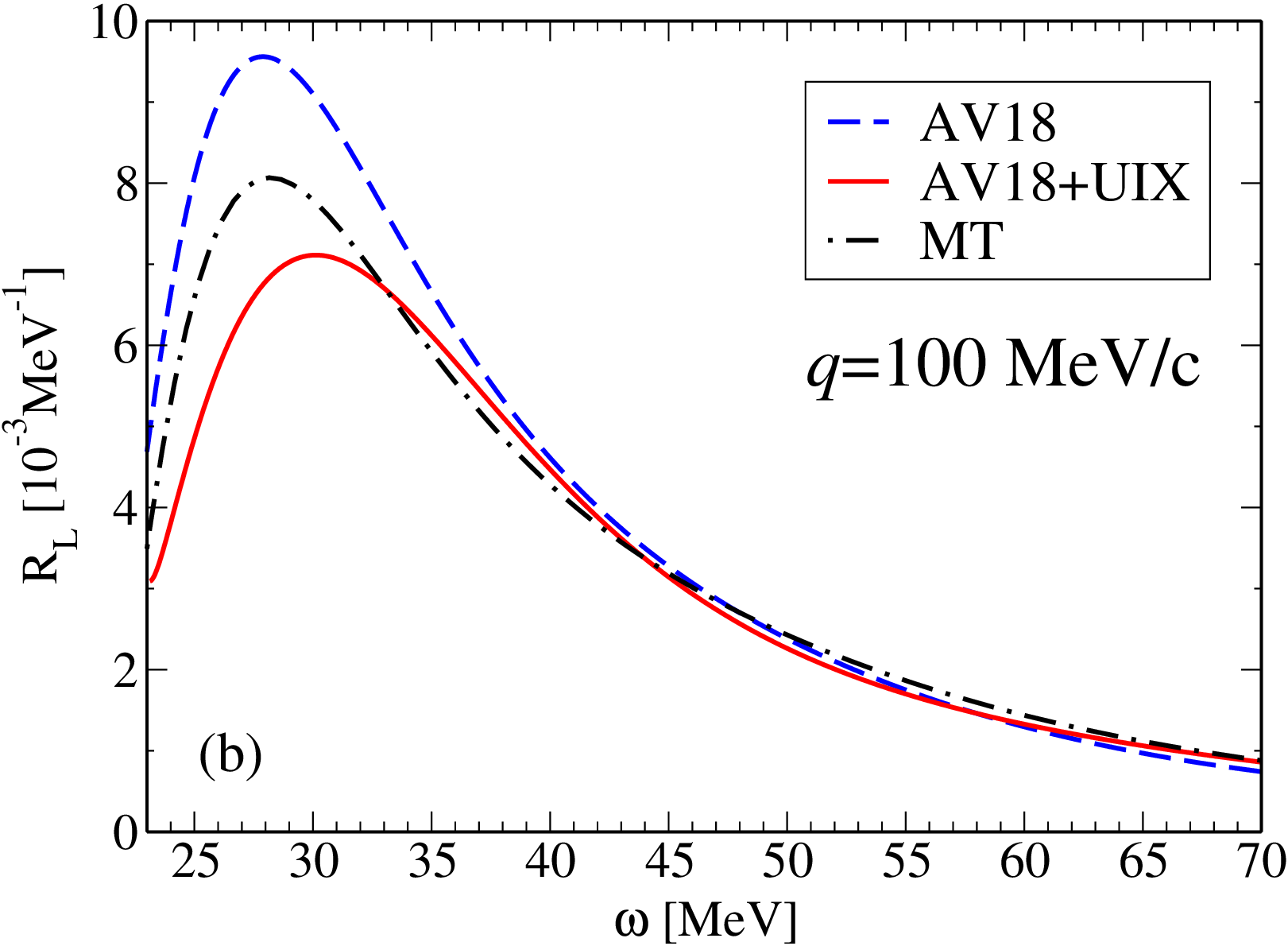}
\caption{(Color online) Longitudinal response function for $q=200$  and $100$ ${\rm MeV}/c$  
with the AV18 (dashed), AV18+UIX (solid) and MT (dashed-dotted) potentials.
Data at $q \simeq 200$  ${\rm MeV}/c$  from \cite{Buki06}.} 
\label{FIG1}
\end{figure}
\begin{figure}
\centering
\includegraphics[scale=0.25,clip=]{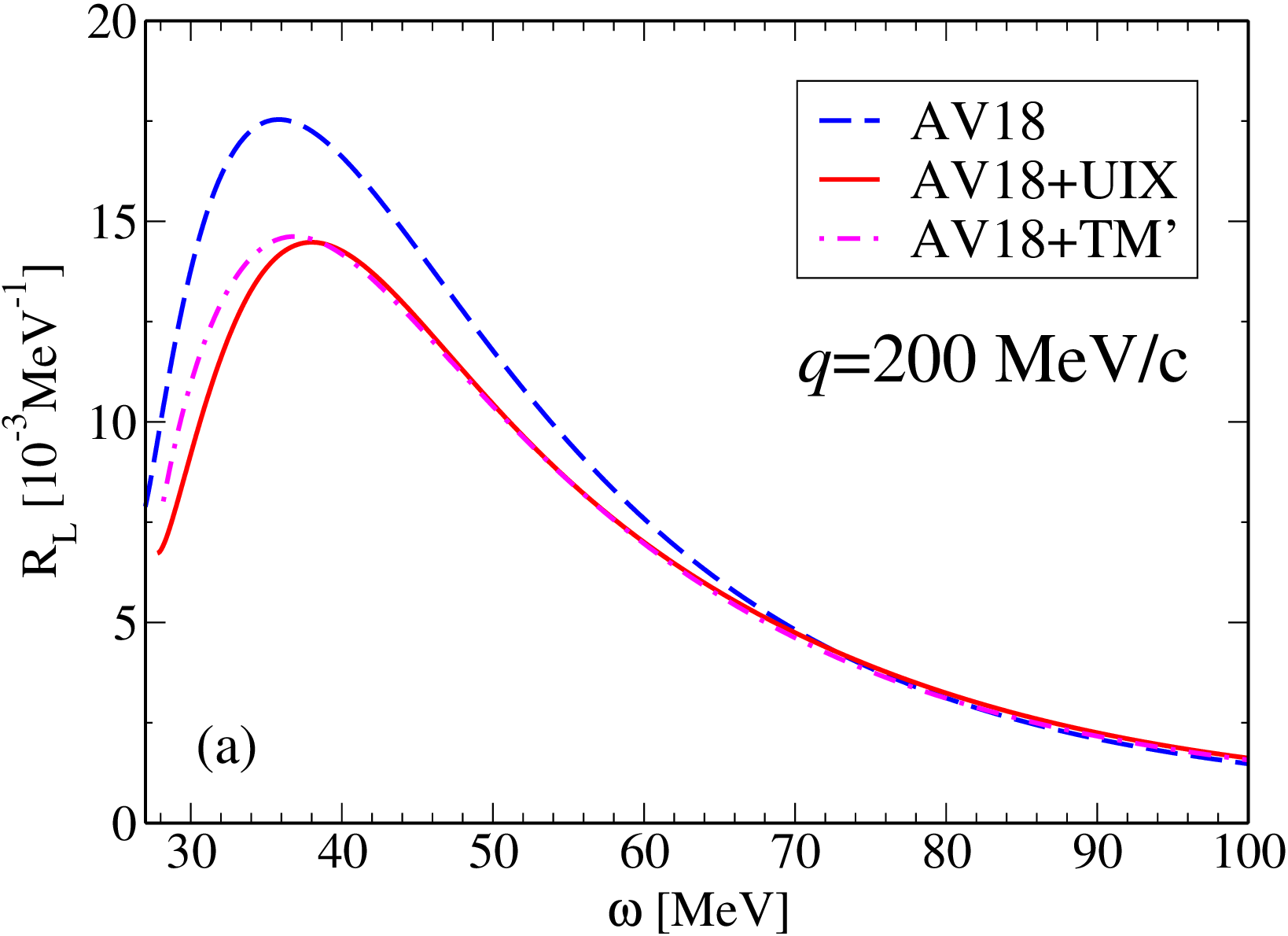}
\\
~\\
\includegraphics[scale=0.25,clip=]{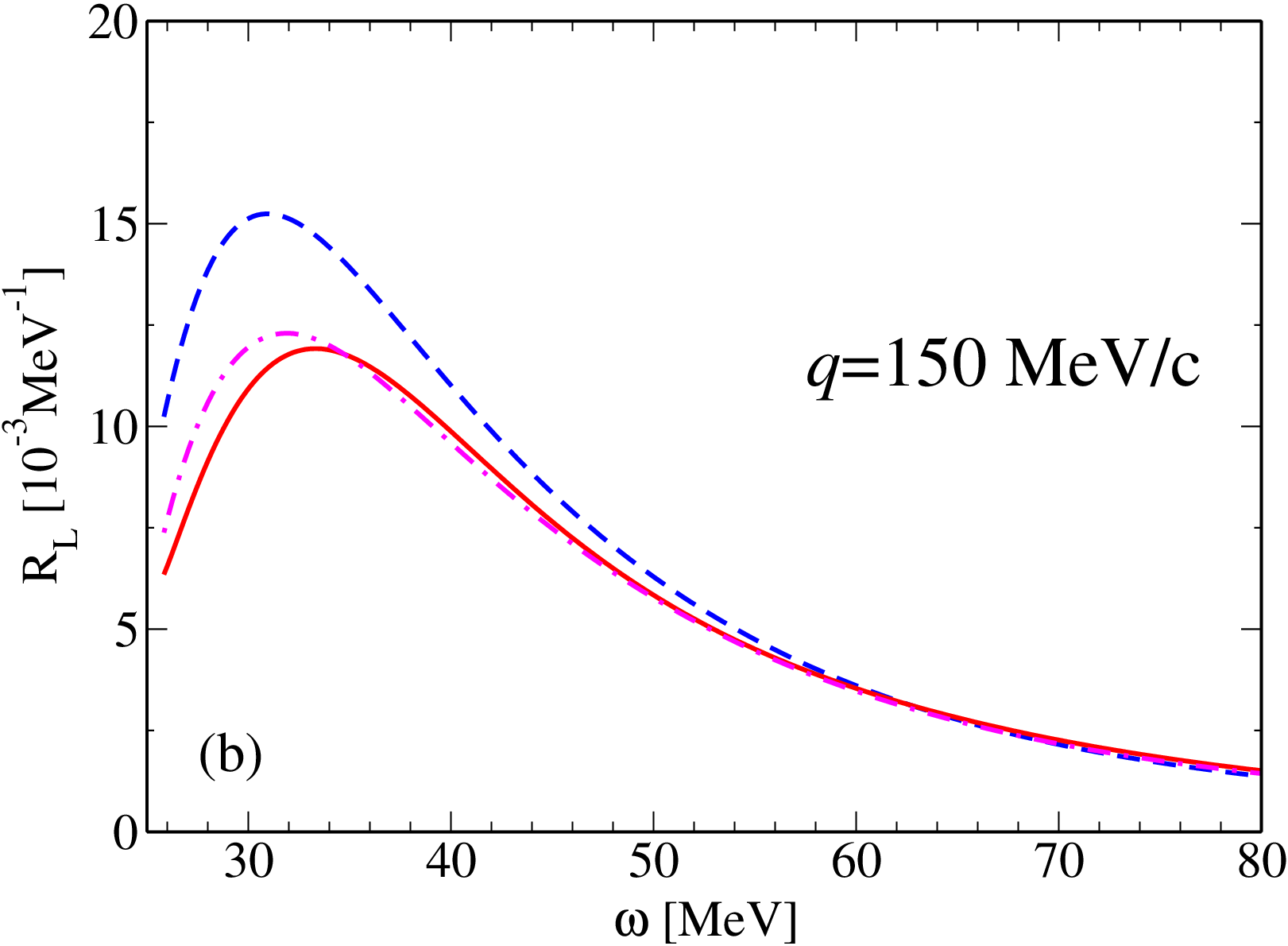}
\\
~\\
\includegraphics[scale=0.25,clip=]{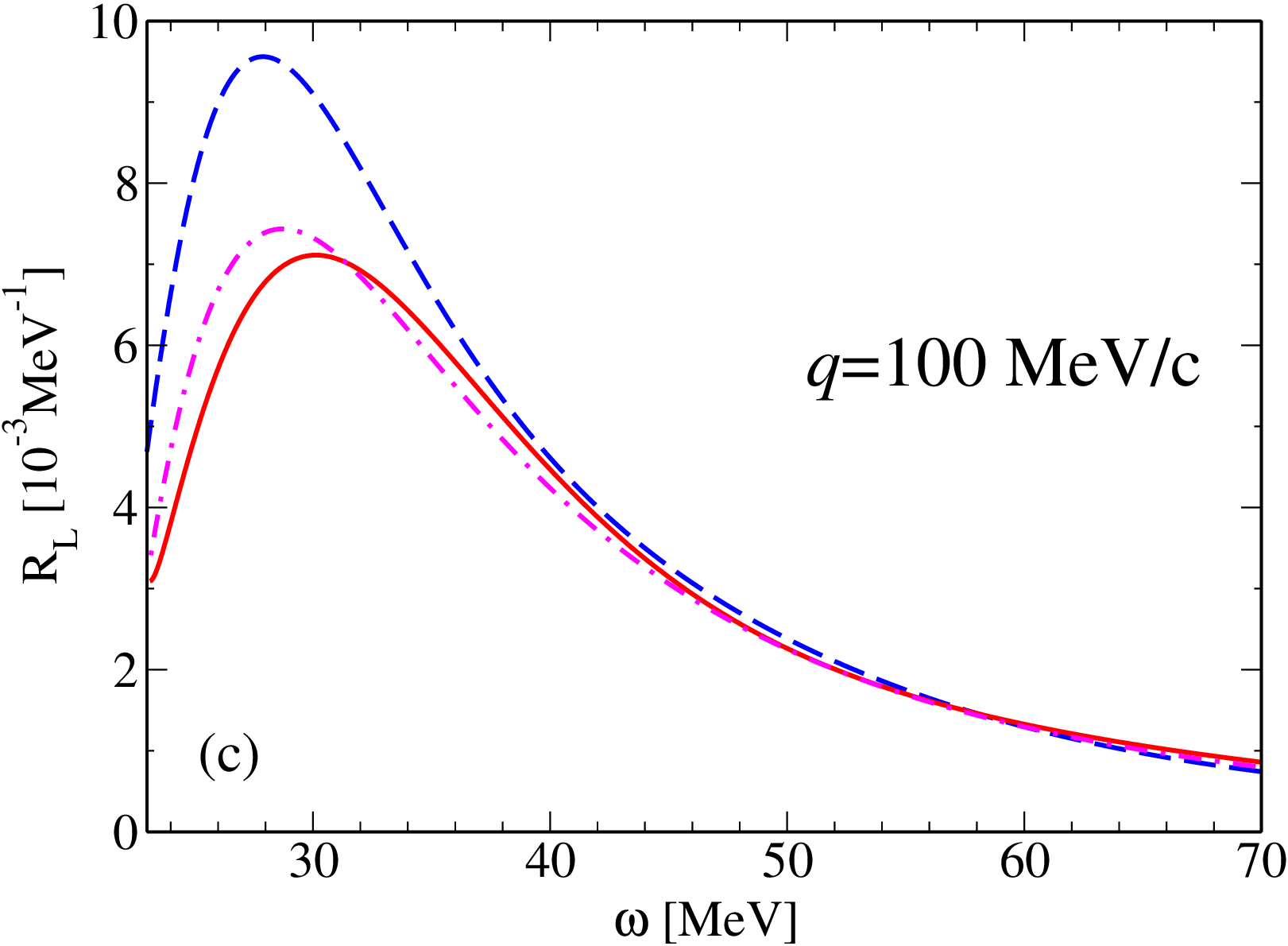}
\\
~\\
\includegraphics[scale=0.25,clip=]{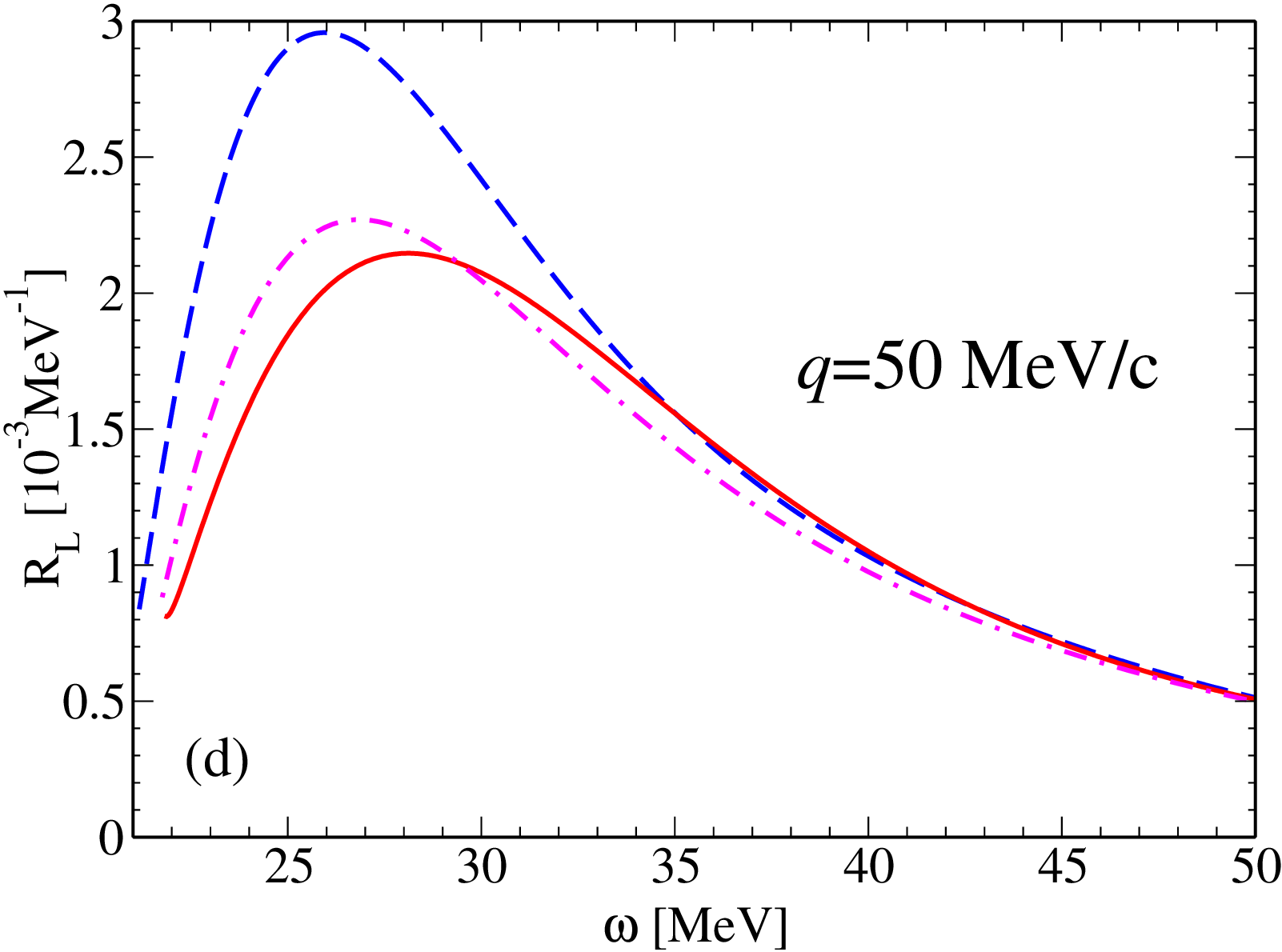}
\caption{(Color online) Longitudinal response function for $q=200,150,100$ and $50$ ${\rm MeV}/c$ 
 with  the AV18 two-nucleon force only (dashed),  and with the addition of the UIX (solid) or the TM' (dashed-dotted) three-nucleon force.}
\label{FIG2}
\end{figure}

\begin{figure*}
\includegraphics[scale=0.25,clip=]{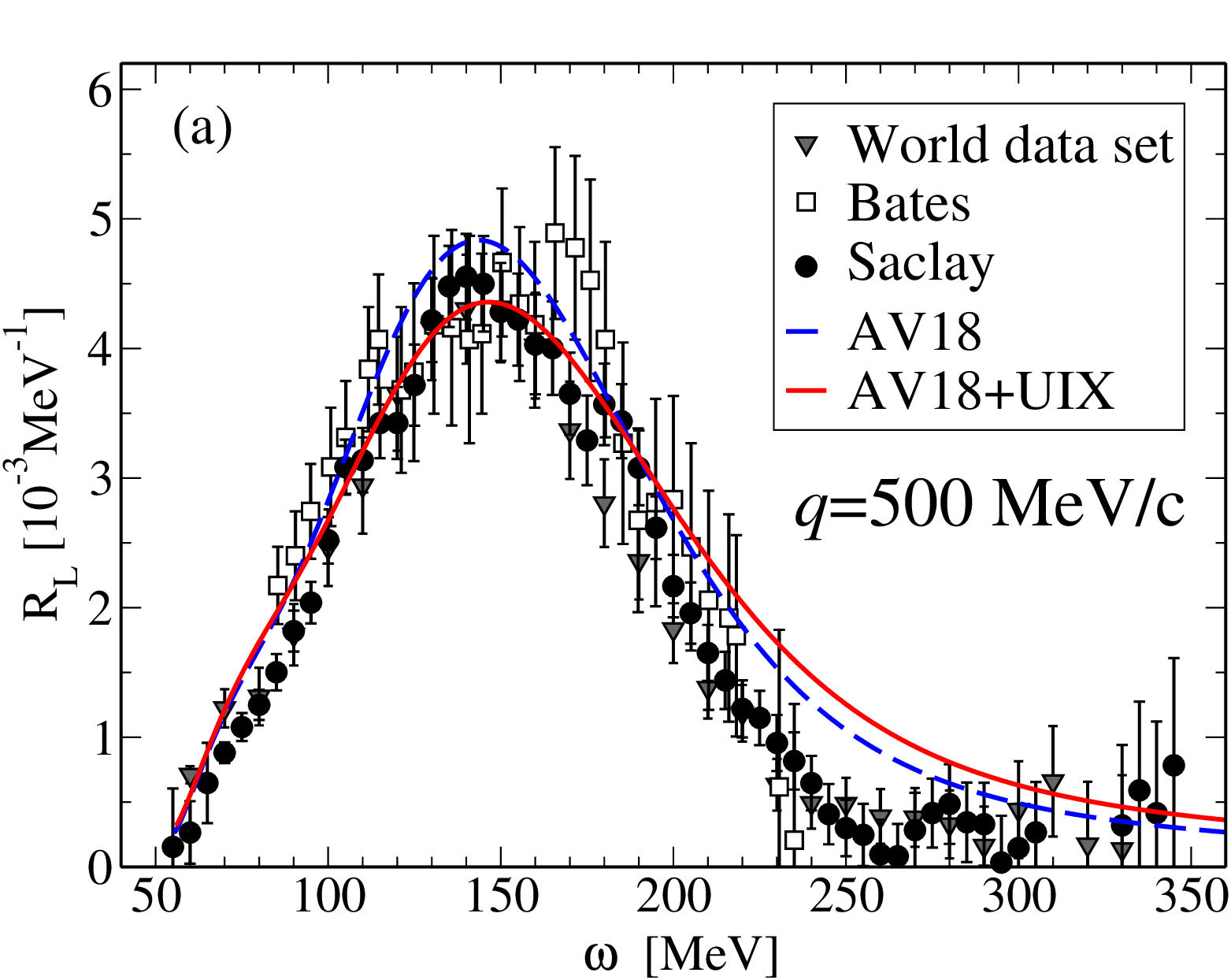}
~~\includegraphics[scale=0.25,clip=]{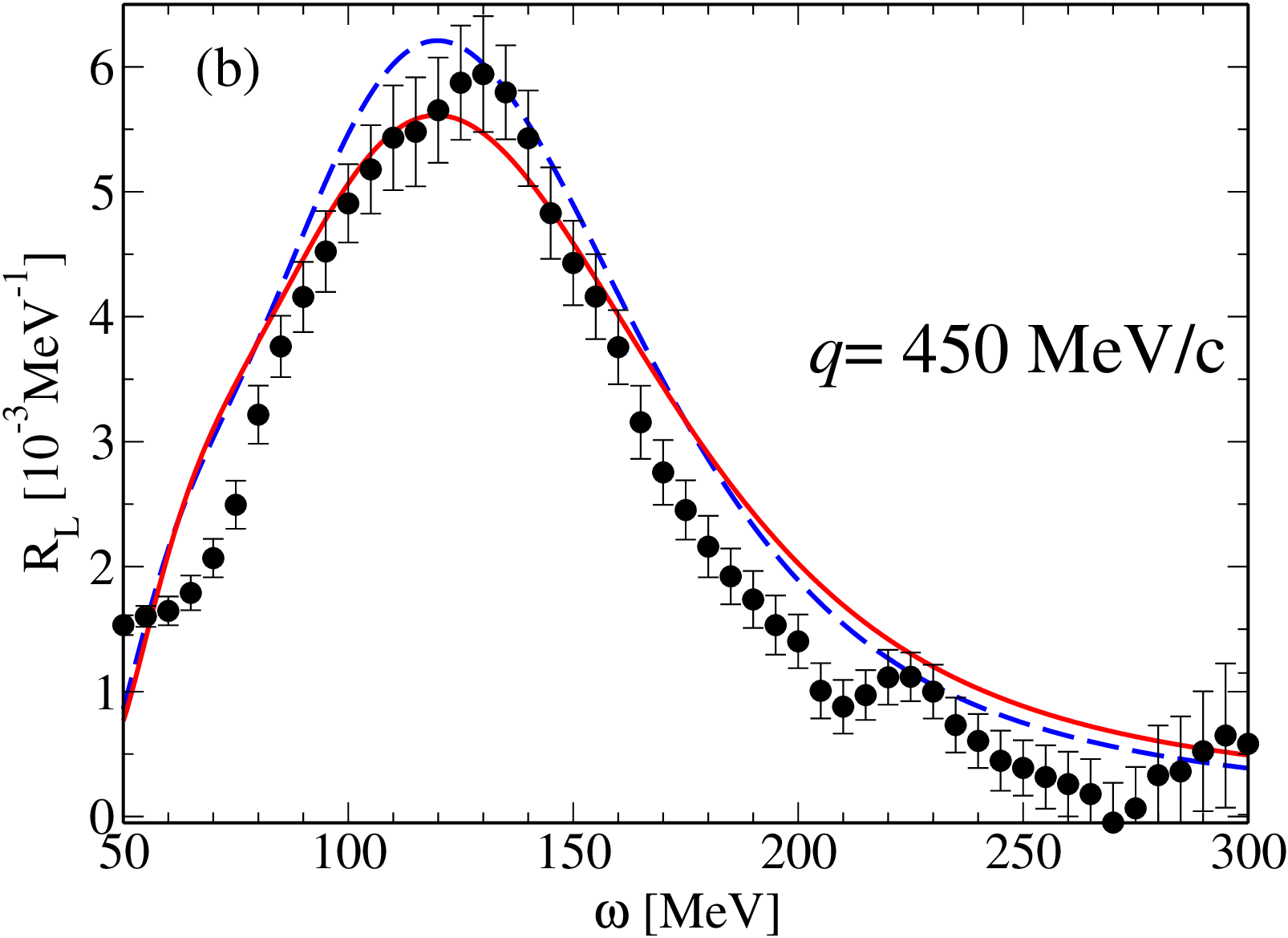}
\includegraphics[scale=0.25,clip=]{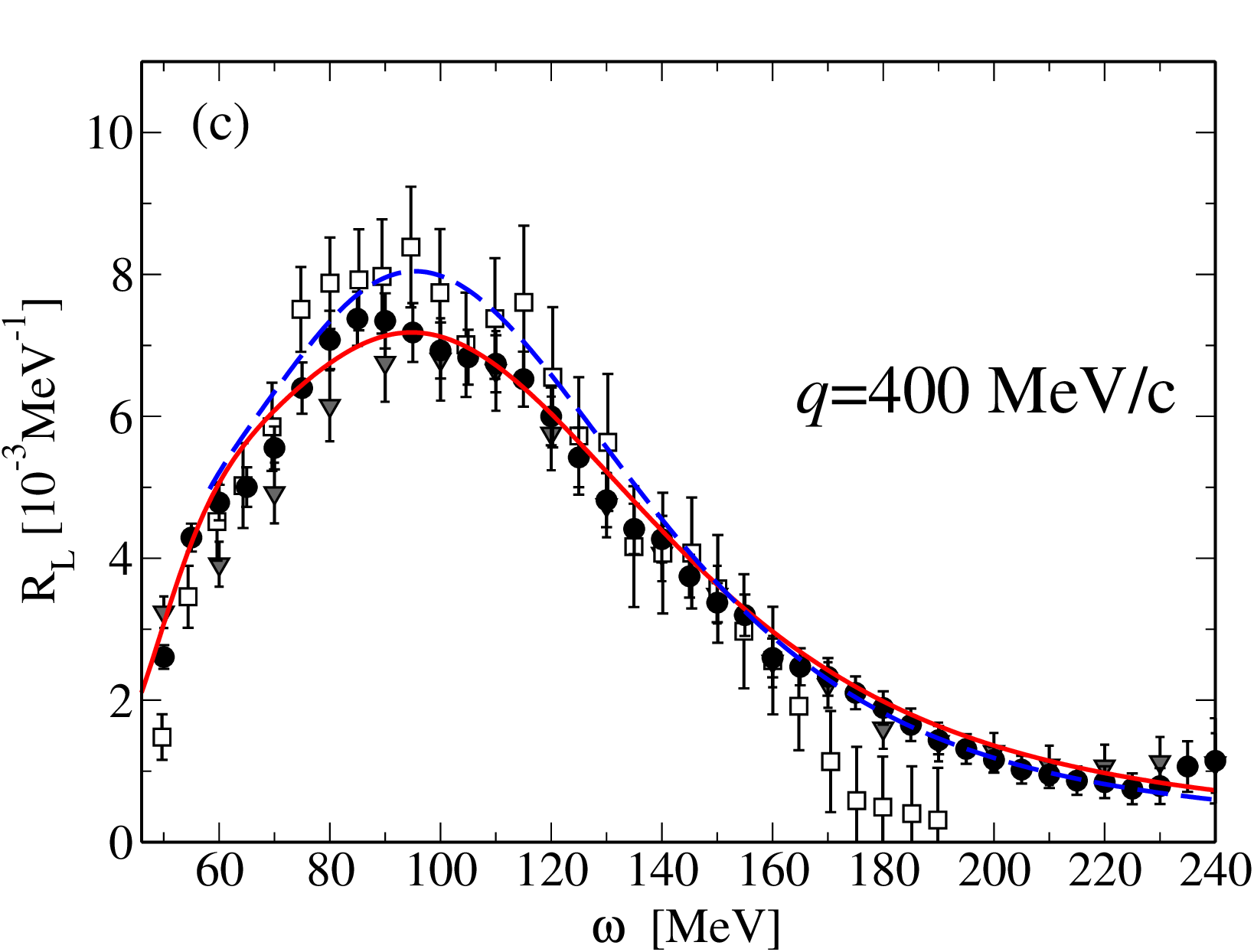}
~~\includegraphics[scale=0.25,clip=]{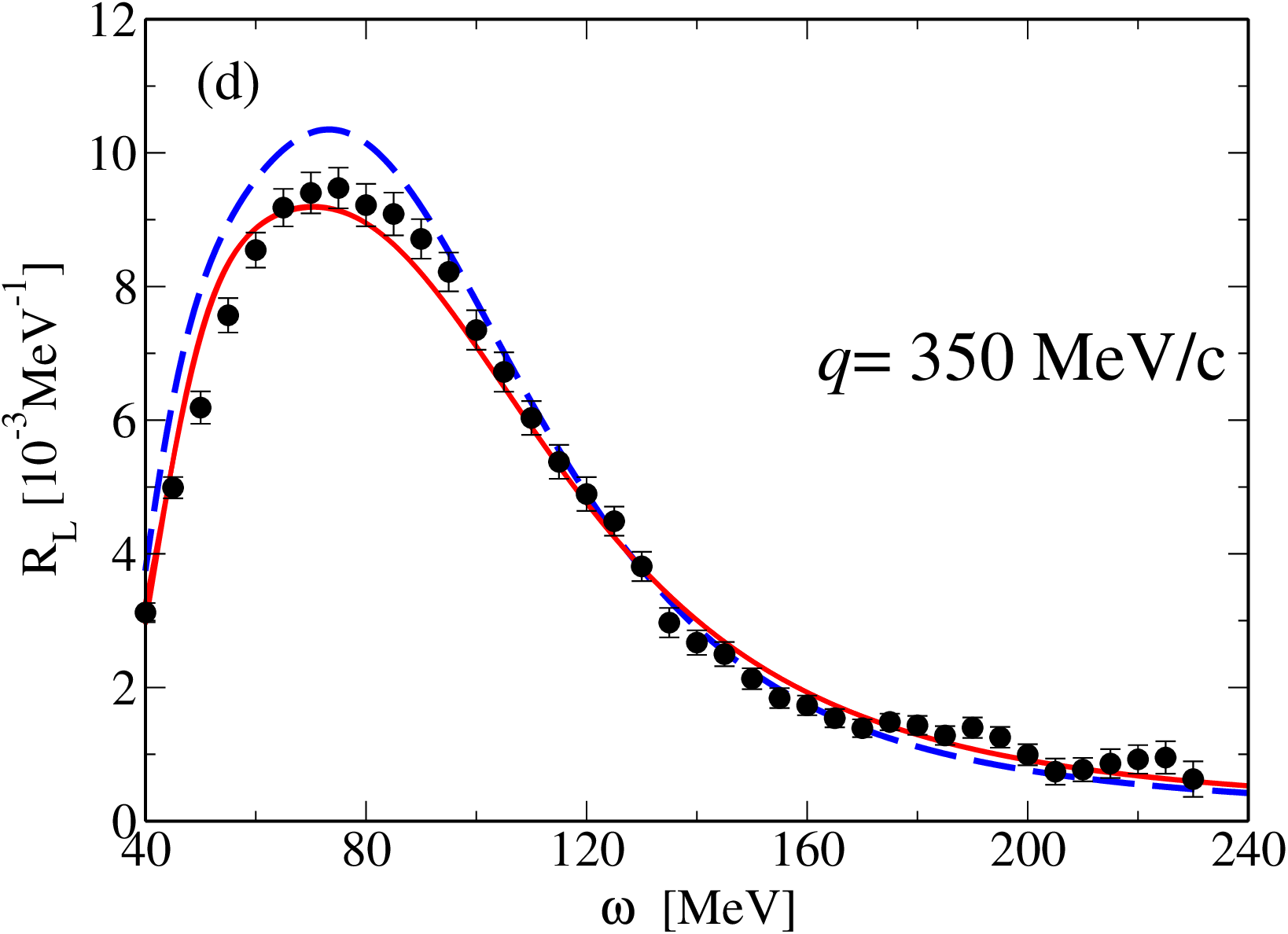}
\includegraphics[scale=0.25,clip=]{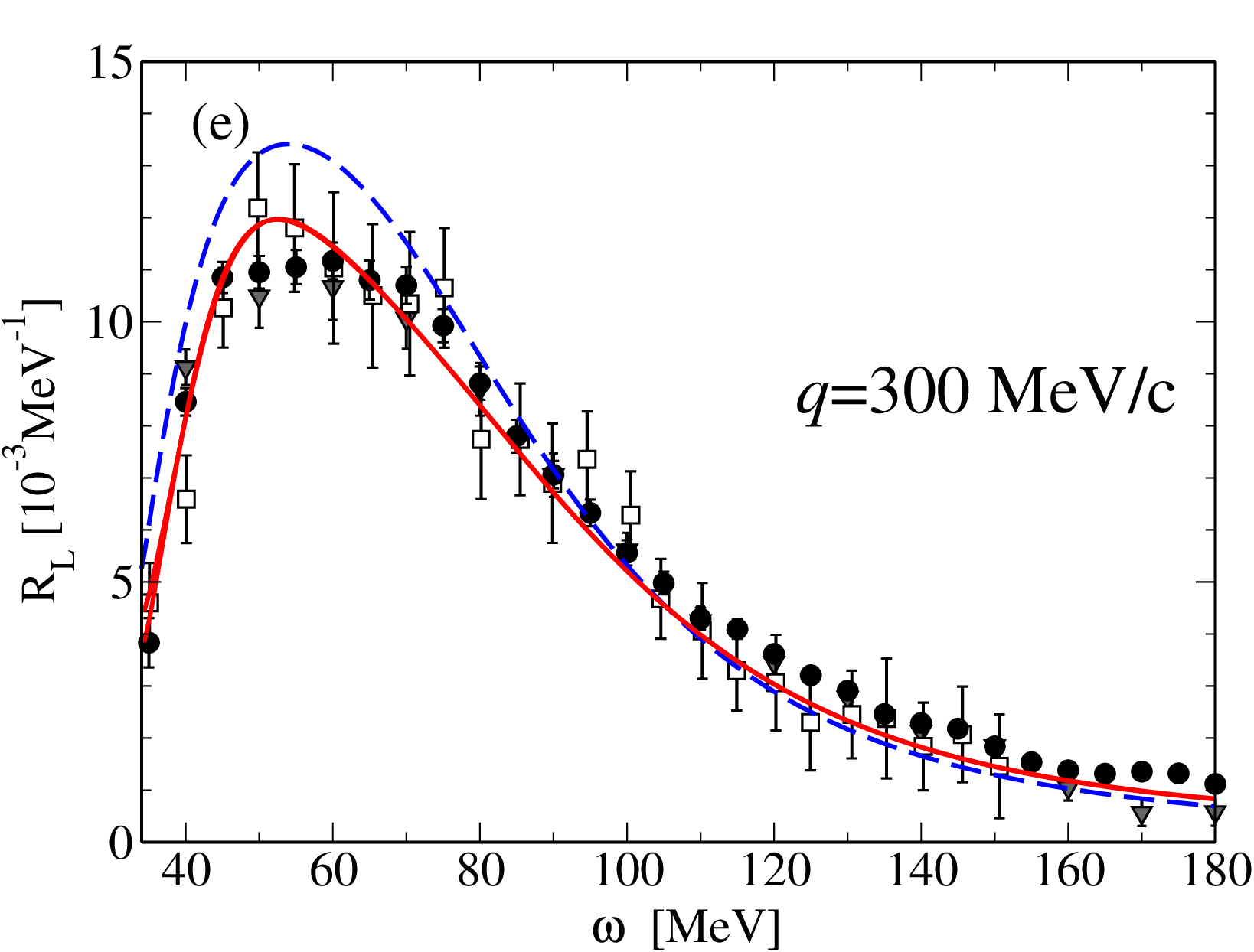}
~~\includegraphics[scale=0.25,clip=]{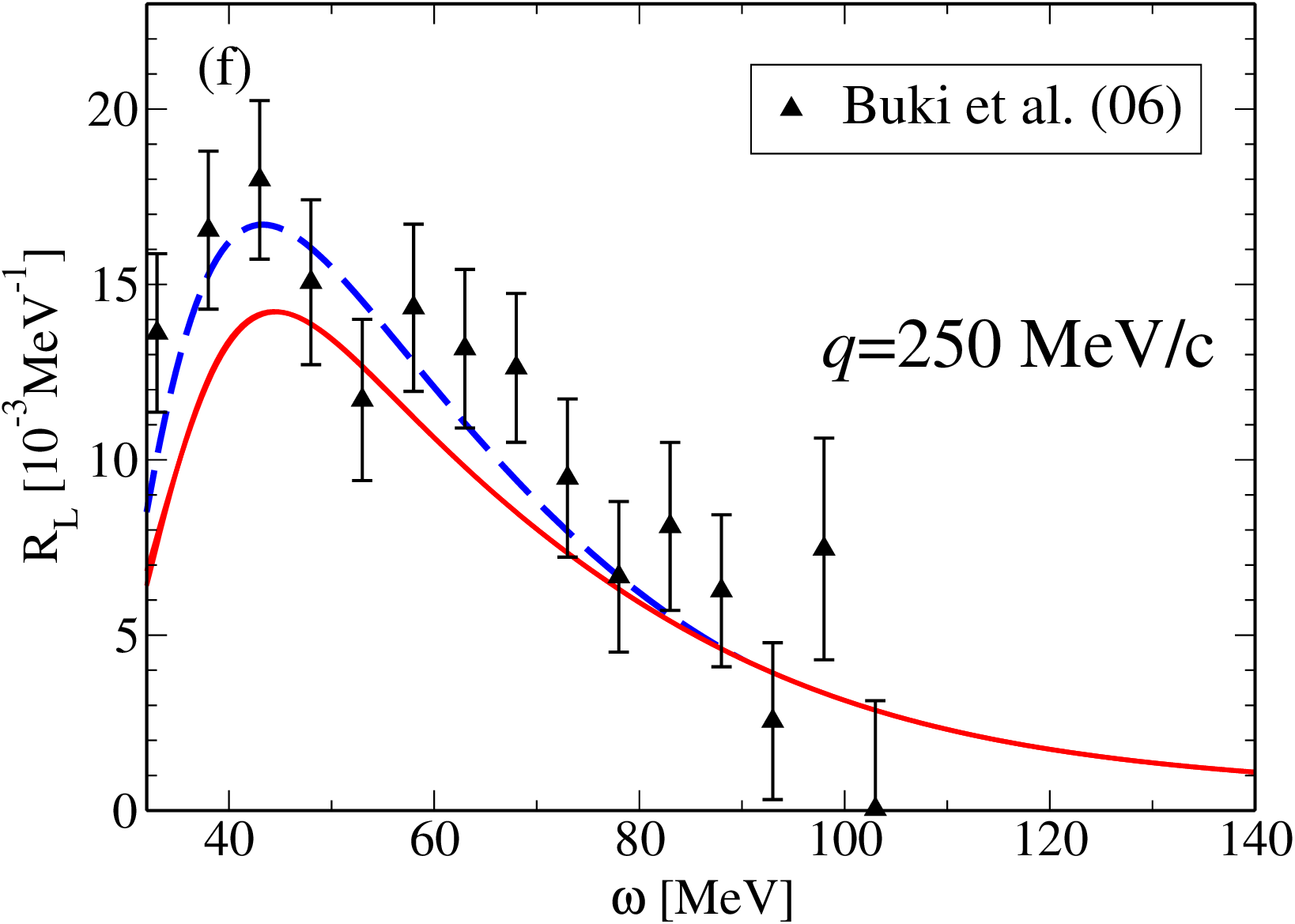}
\caption{(Color online) $R_L(\omega,q)$ for $250 \le q \le 500$ ${\rm MeV}/c$:  
calculations with AV18 (dashed) and AV18+UIX
(solid). Data from~\cite{Bates} (squares),~\cite{Saclay} (circles),~\cite{Carlson02} (triangles down), and~\cite{Buki06}
(triangles up).} 
\label{FIG3}
\end{figure*}

In this section we  present results on $R_L$, focusing on the evolution 
of dynamical effects as the momentum transfer decreases. In Fig.~\ref{FIG1}, we show 
$R_L$ at constant $q= 200$ and 100 MeV/c.
As already shown in~\cite{PRL09} one has a large quenching effect due the 3NF, which 
is strongest at lower  $q$. 
One should notice that such an effect is not simply correlated to the under-binding of the AV18
potential (binding energy $E_B$= 24.35 MeV in the present LIT calculation, with higher EIHH
precision $E_B$= 24.27 MeV \cite{Gazit06}). In fact, if this was the case, the results with the Malfliet-Tjon potential 
(MT)~\cite{MaT69}, which gives a slight over-binding of $^4$He ($E_B$=30.56 MeV), would lay even 
below those obtained with AV18+UIX ($E_B$= 28.40). 
On the contrary the MT curve is situated
between the curves with and without 3NF. 

We would like to mention that we do not give results for the very threshold region, where one has a 0$^+$ resonance~\cite{Wal70}. 
In our present calculation we are not able to resolve this narrow resonance, therefore we subtract its contribution
before inversion as we do for the elastic peak \cite{REPORT07}. This procedure of course does not affect the result above the resonance.

Given the large 3NF effect at lower $q$ it is interesting to see whether
there is a dependence of the results on the  3NF model itself. To this end
we have performed the calculation using also the Tucson Melbourne (TM')~\cite{TMprime}  three-nucleon force. 
While the UIX force contains a two-pion exchange and a short range phenomenological term, with two 3NF parameters fitted
on the triton binding energy and on nuclear matter density (in conjunction with the AV18 two-nucleon potential), the TM' force
is not adjusted in this way. It includes two pion exchange terms where the coupling constants are taken from pion-nucleon scattering 
data consistently with chiral symmetry.

Our results with the TM' force are obtained using the same model space as for the UIX potential and
the accuracy of the convergence for the LIT is found to be at a percentage level, in analogy to that of the UIX as
described in \cite{PRL09}.
The cutoff of the TM' force has been adjusted on triton binding energy, when used in conjunction with the AV18 NN force.
With a cutoff $\Lambda=4.77m_\pi$, where $m_\pi$ is the pion mass, we obtain the following binding energies
8.47 MeV ($^3$H) and 28.46 MeV ($^4$He). We  would like to emphasize that the $^4$He binding energy 
is practically the same as for the AV18+UIX case (as already found in \cite{Nogga02}).

In addition to what is shown in our recent Letter \cite{PRL09}, here we investigate also other
low-$q$ values. Figure~\ref{FIG2} shows that the increase of 3NF effects 
with decreasing $q$ is confirmed. Moreover
it becomes evident that also the difference between the results obtained with two  3NF models 
increases  with decreasing $q$. One actually finds that
the shift of the peak to higher energies in the case of UIX  generates for $R_L$ a difference up to about 10\%
on the left hand sides of the peaks. This is a very interesting result. It represents the first case of an 
electromagnetic observable considerably dependent on the choice of the 3NF.
In the light of these results it would be very interesting to repeat the calculation with  EFT 
two-and three-body potentials \cite{EM,EGM}. At the same time it would be highly desirable 
to have precise measurements of $R_L$ at low $q$. This could serve either to fix the 
low-energy constants (LEC) of the effective field theory 3NF or to possibly discriminate between different
nuclear force models.

In Fig.~\ref{FIG3}, an overview of the results obtained for larger $q$ is given, showing also 
the comparison with existing experimental data. One sees that the 3NF results are closer to the data, this 
is particularly evident at $q=300$ MeV/c. However, the 3NF effect is generally not as large as for the lower momentum transfers shown
in Figs.~\ref{FIG2} and~\ref{FIG3}. In some cases the quenching of the strength due to the 3NF is  
comparable to the size of the error bars, particularly for the data from Ref.~\cite{Bates}.
The largest discrepancies with data are found at $q=500$ MeV/c.
While the height of the peak is well reproduced by the result with 3NF, the width
of the experimental peak seems to be somewhat narrower than the theoretical one. On the other hand
 one has to be aware that relativistic effects
are not completely negligible at $q=500$ MeV/c. They probably play a similar role as found in the electro-disintegration
of the three-nucleon systems (see e.g.  Ref.~\cite{RL3B}). In the case of $q=$ 250 MeV/c
the experimental results are not sufficiently precise to draw a conclusion.

\section{Multipole Analysis}\label{MA} 

\begin{figure*}
\includegraphics[scale=0.60,clip=]{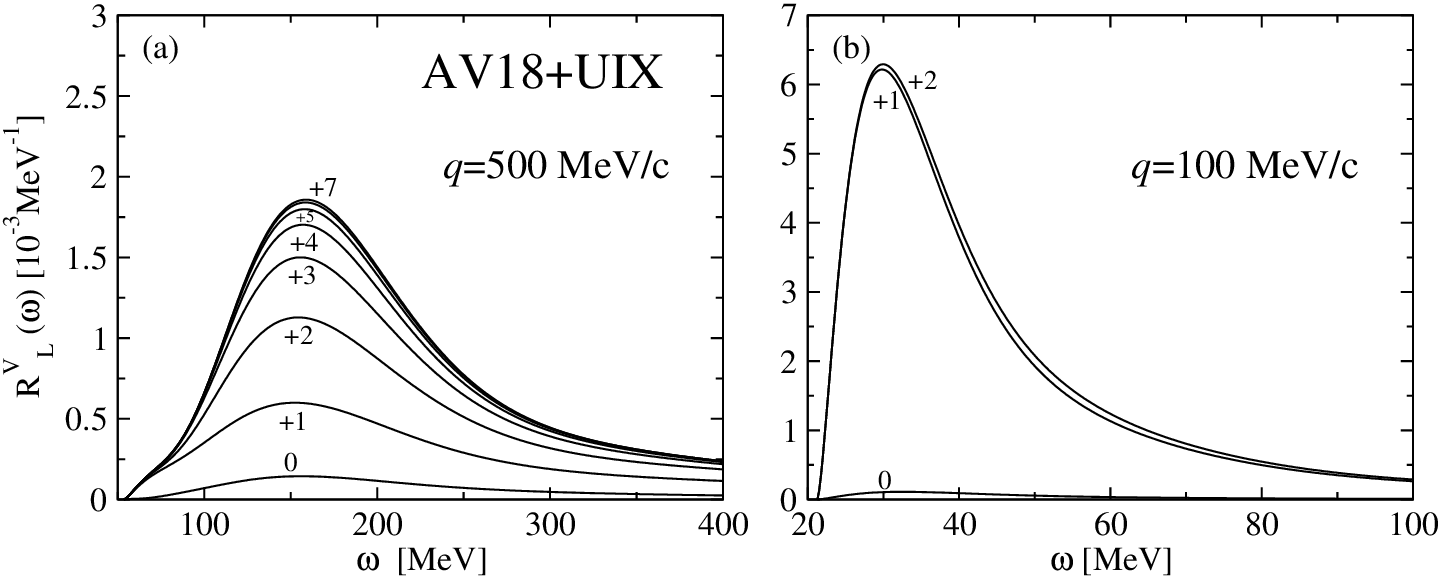}\\
~\\
\includegraphics[scale=0.60,clip=]{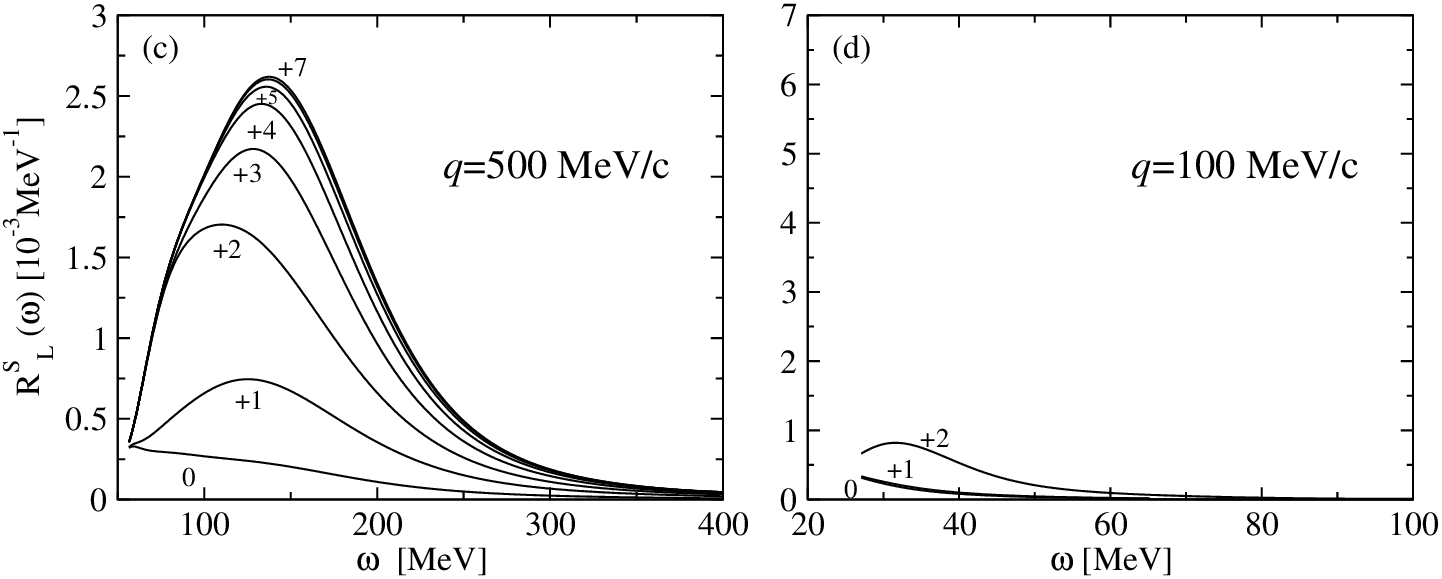}
\caption{Response functions of the lowest isovector (upper panel) and isoscalar (lower panel) Coulomb multipoles,
starting with the monopole and consecutively adding higher multipoles
up to $J_{max}=7$ for ${q}=500$~MeV/c (left) and $J_{max}=2$ for
${q}=100$~MeV/c (right) in case of the AV18+UIX potential.}
\label{FIG4}
\end{figure*}

\begin{figure*}
\includegraphics[scale=0.25,clip=]{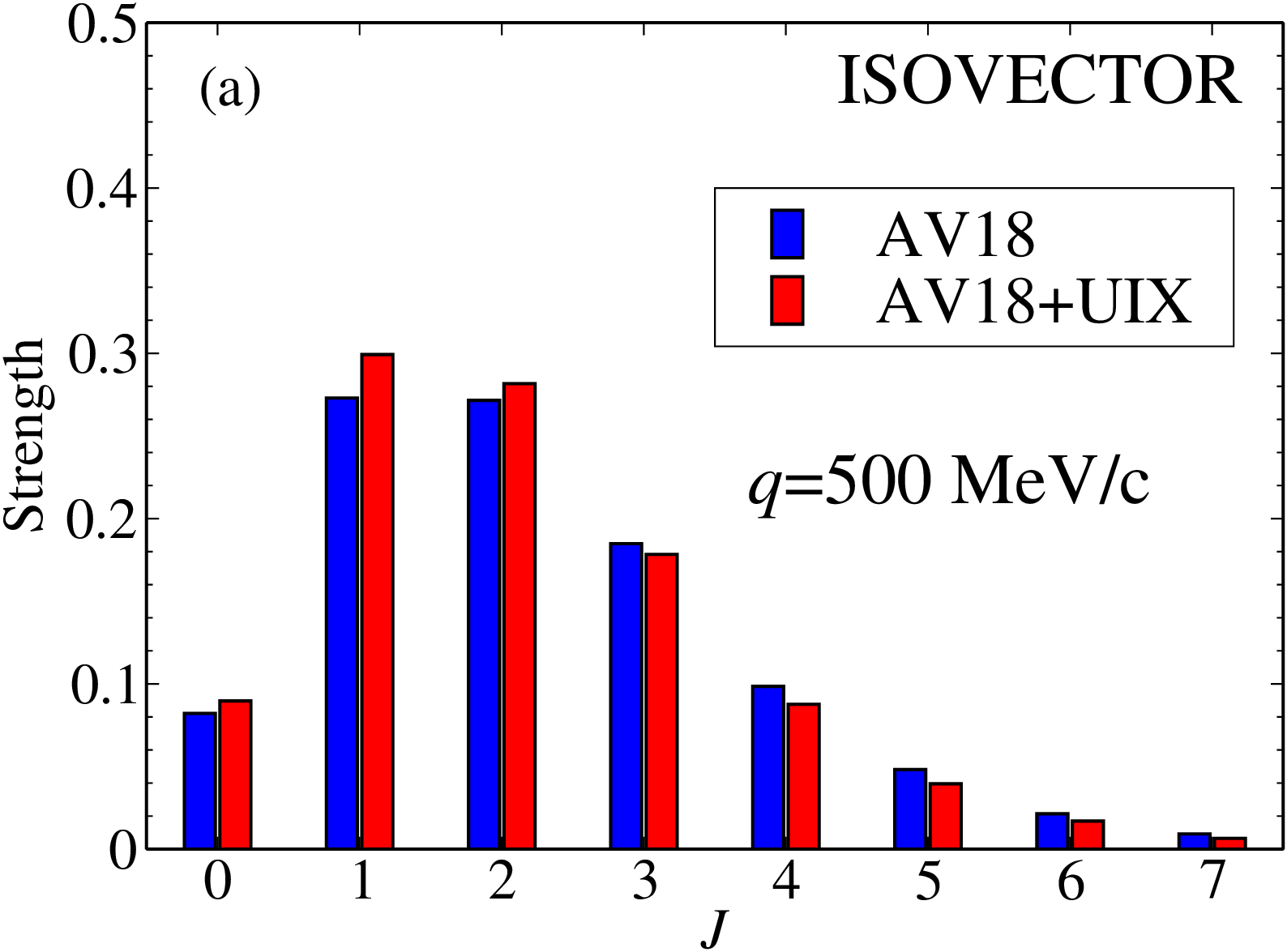}
\includegraphics[scale=0.25,clip=]{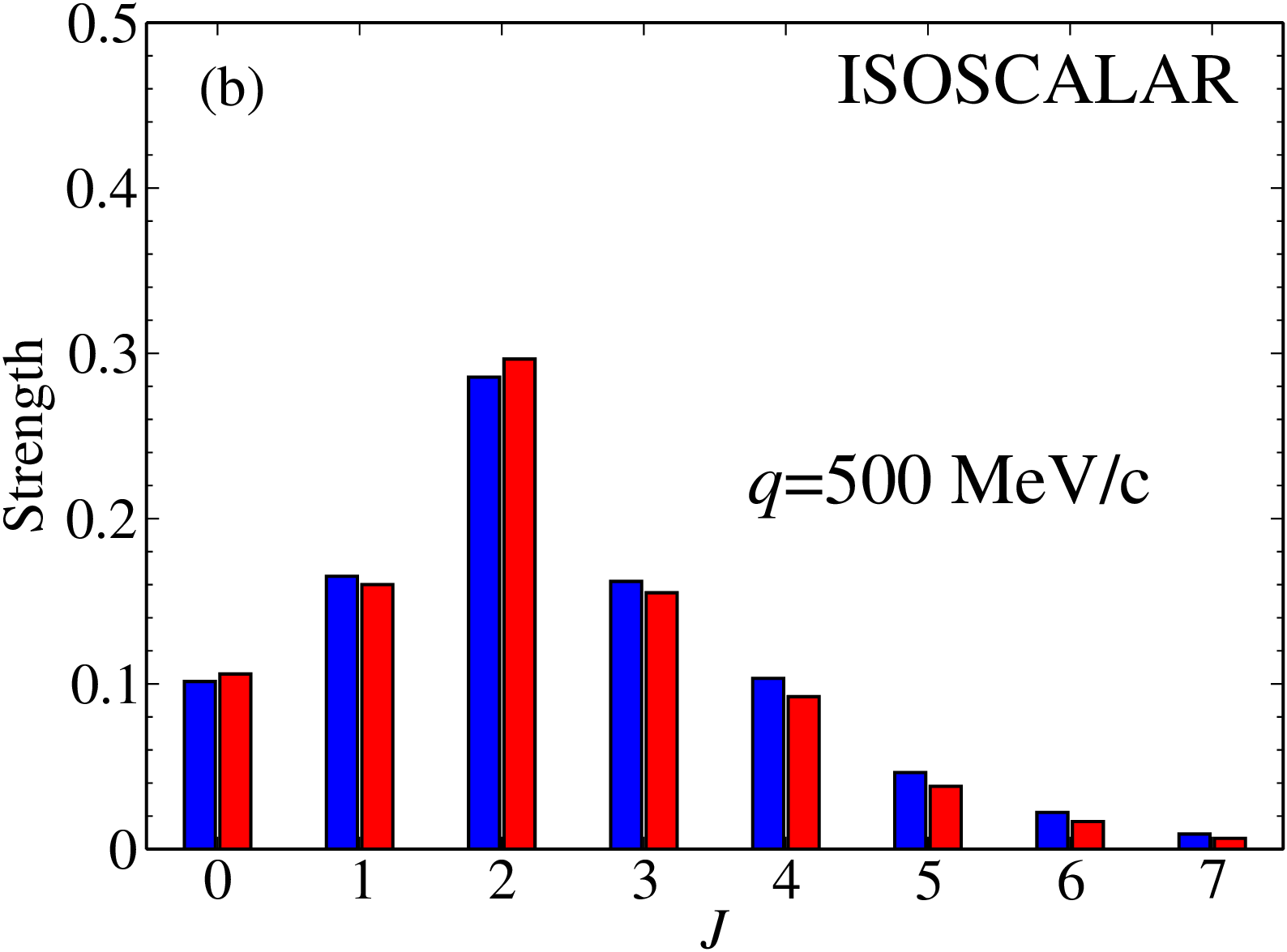}
\includegraphics[scale=0.25,clip=]{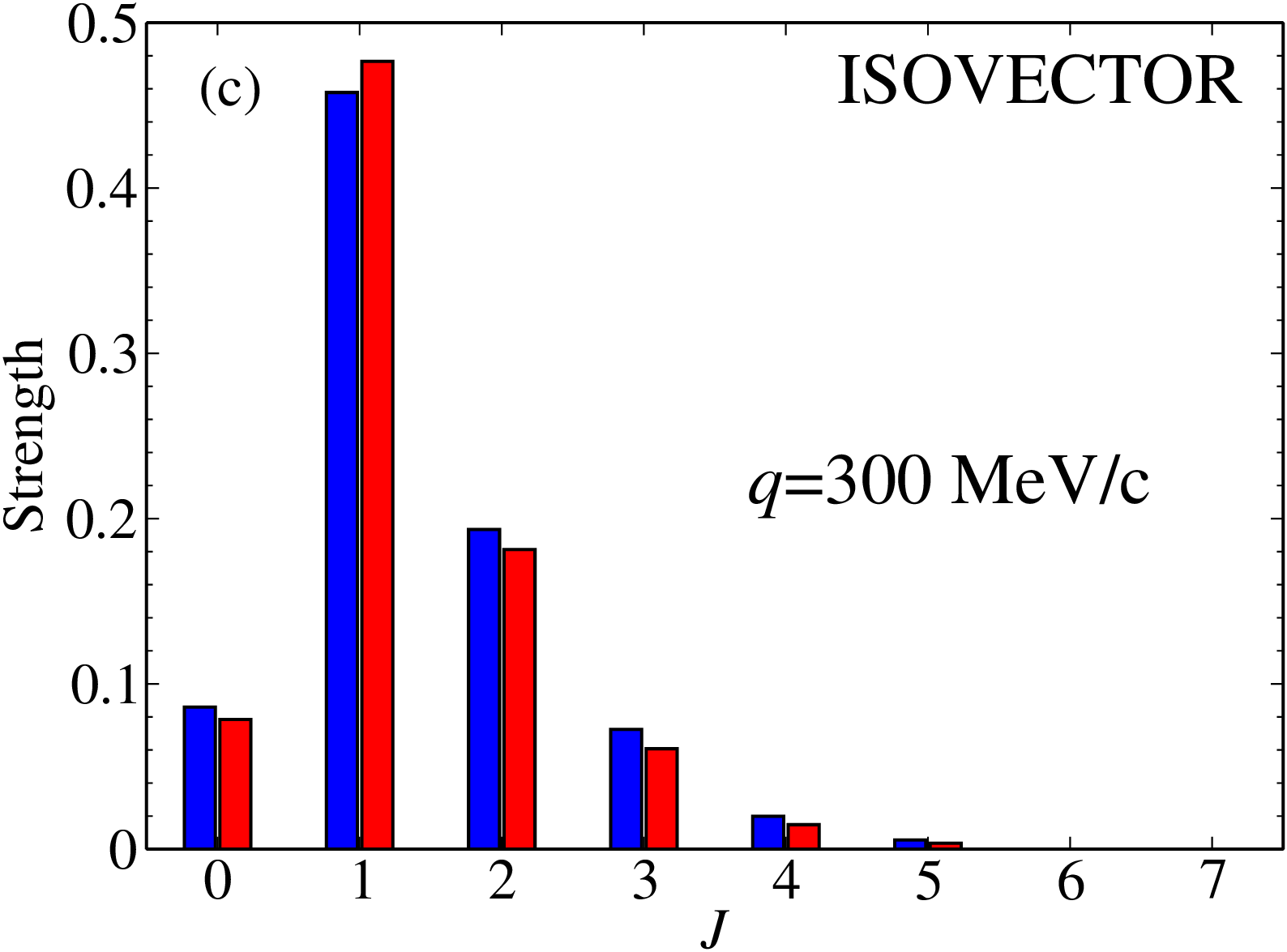}
\includegraphics[scale=0.25,clip=]{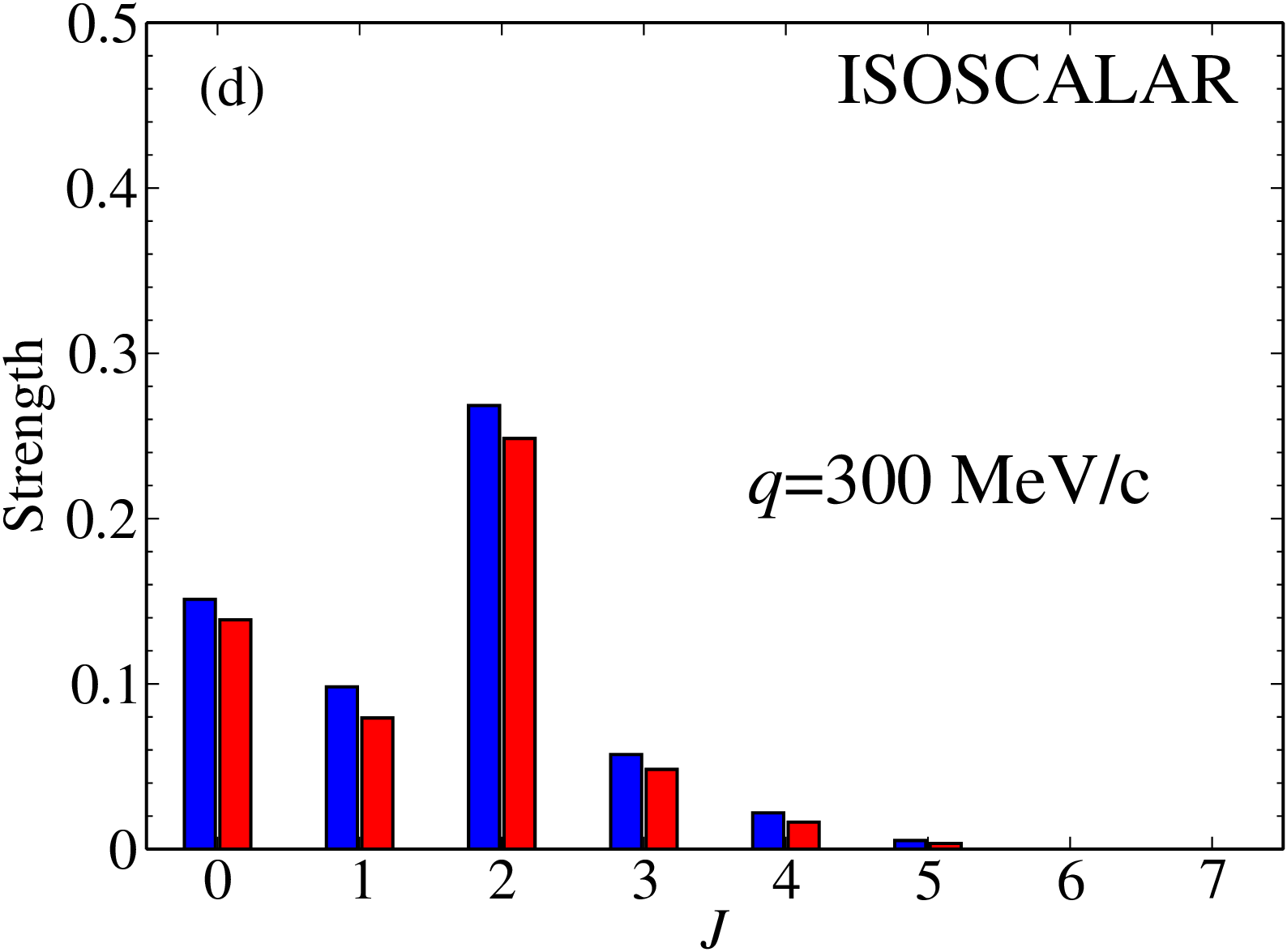}
\includegraphics[scale=0.25,clip=]{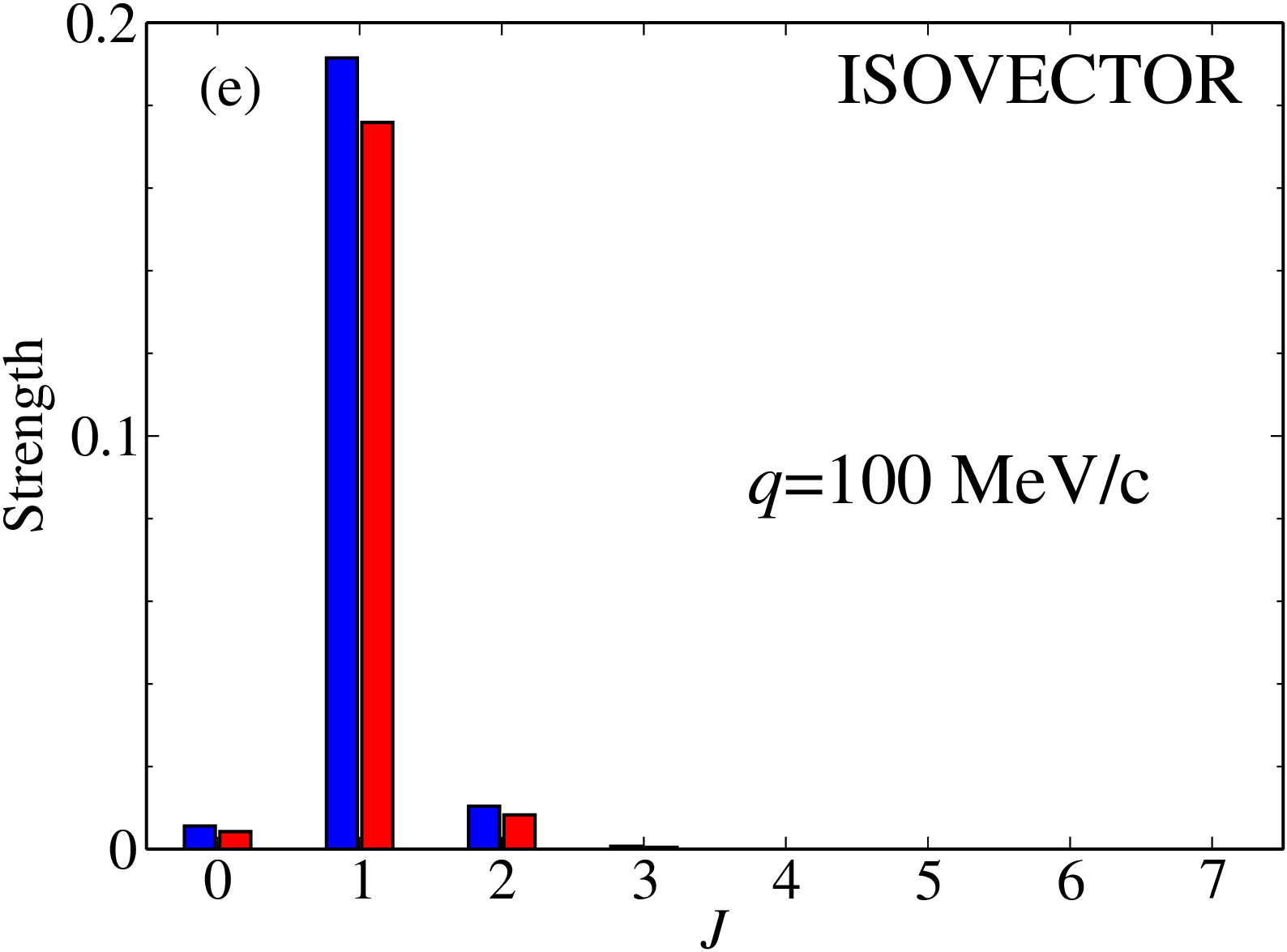}
\includegraphics[scale=0.25,clip=]{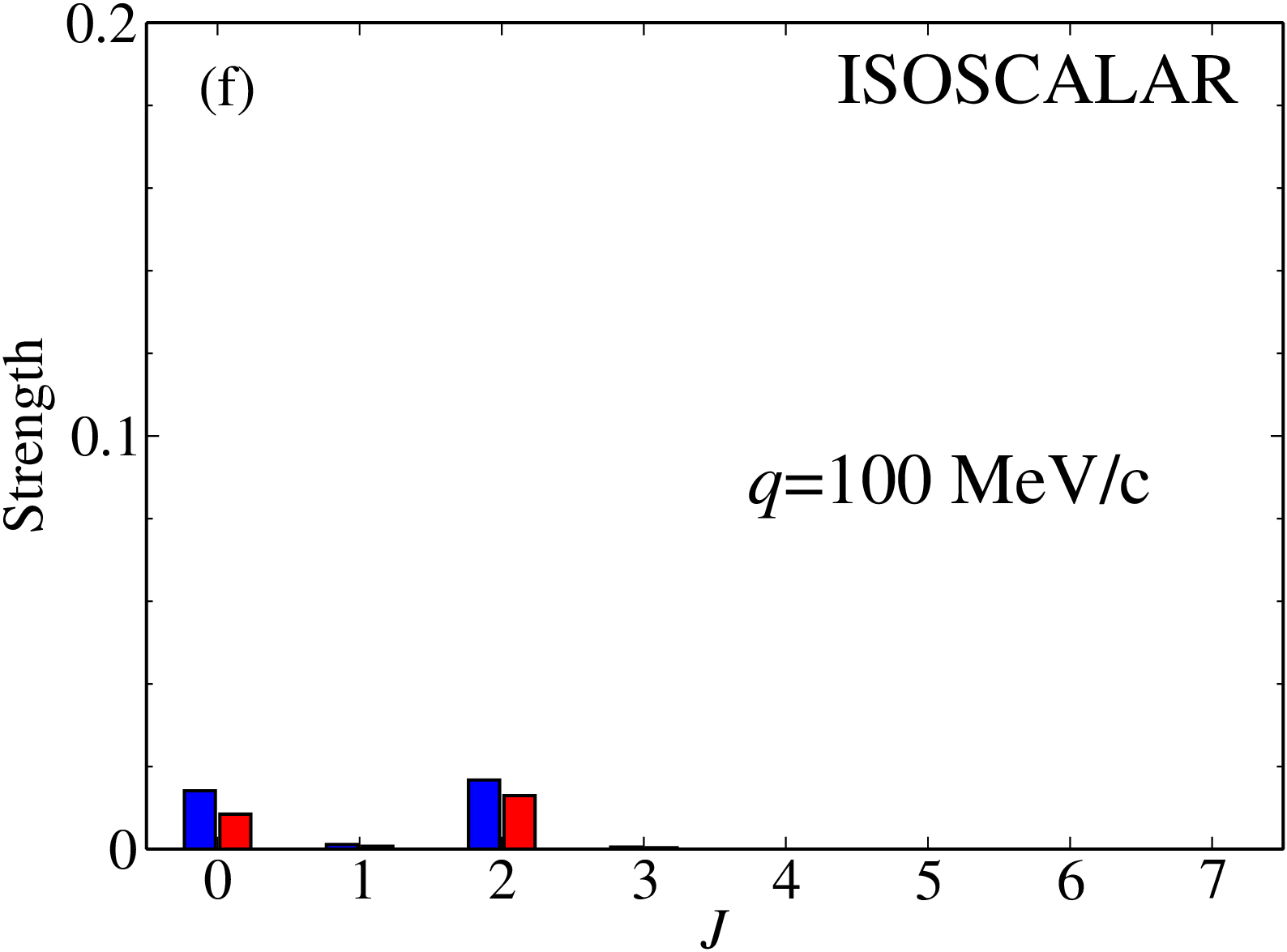}
\caption{(Color online) Isovector (left) and isoscalar (right) multipole strength 
distribution of $R_L$ at $q$=500, 300 and 100 MeV/c in case of the AV18 and 
AV18+UIX potentials. \\
} 
\label{FIG5}
\end{figure*}

It is interesting to analyze the results of $R_L$ in terms of its multipole contributions.
Using Eq.~(\ref{mult_charge}) on 
the right-hand-side of Eq.~(\ref{liteq}), one can decompose  
${\cal L}_L(\sigma,q)$ into a sum of multipole contributions ${\cal L}^{J,X}(\sigma,q)$.
We have calculated each of them separately, solving the corresponding equations~(\ref{liteq}). 
After the inversion of the transform we have obtained the various multipole responses. By multiplying them 
with the isoscalar/isovector nucleon form factors we generate 
the multipole contributions  to  the longitudinal response function   $R_L^{J,X}(\omega,q)$. 
In Fig.~\ref{FIG4} we show how the isoscalar and isovector parts of $R_L$ are built up from their 
multipole contributions at a higher (500 MeV/c)  and a lower (100 MeV/c) value of $q$.
As expected, the higher the momentum transfer, the larger the number of multipoles that have to be 
considered to reach convergence. For $q=500$ MeV/c up to seven multipoles are considered, while for $q=100$ MeV/c
only three multipoles are required for a converged result.
Regarding the strength distribution among the multipoles two facts are evident: 
i) at higher $q$ the strength is almost equally distributed  among the first isovector 
multipoles, while in the isoscalar channel the quadrupole gives the largest contribution;
ii) at low $q$, as expected, the response is dominated by the isovector dipole contribution. While 
the isoscalar dipole  is completely  negligible, the isoscalar quadrupole contributes a 
few percent.
A careful reader may notice that the isoscalar response at $q=100$ MeV/c does not seem to show a 
convergence in the multipole decomposition. As already mentioned, the isoscalar dipole is negligible
(explaining why the curve labeled with ``0" is overlapping with the ``+1" one), and a similarly negligible
strength is found for the multipoles higher than the quadrupole. This fact is also seen in Fig.~\ref{FIG5}, where
the total strength of the various multipoles is shown for 
$q=100$, 300 and 500 MeV/c.
\begin{figure*}
\includegraphics[scale=0.25,clip=]{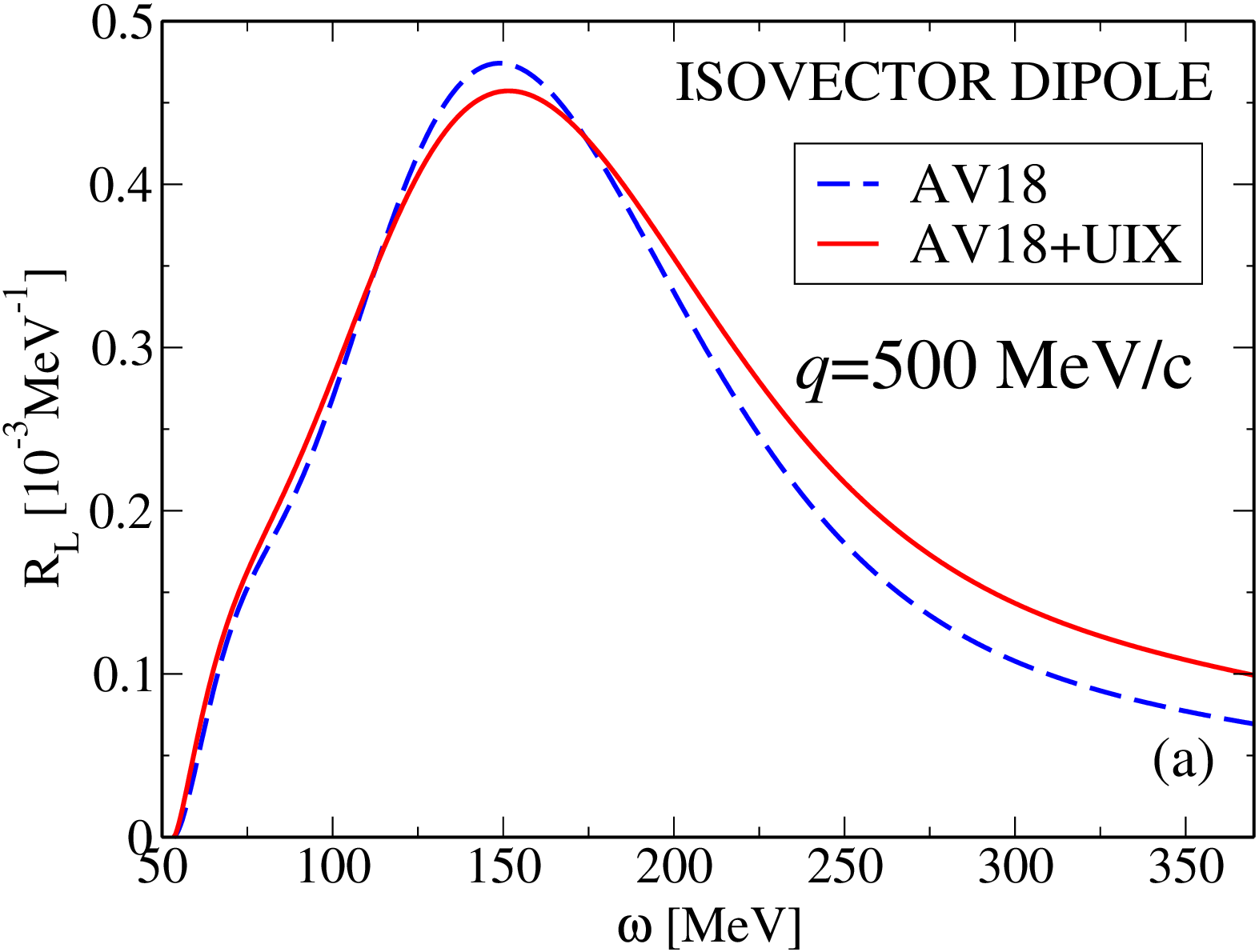}
\includegraphics[scale=0.25,clip=]{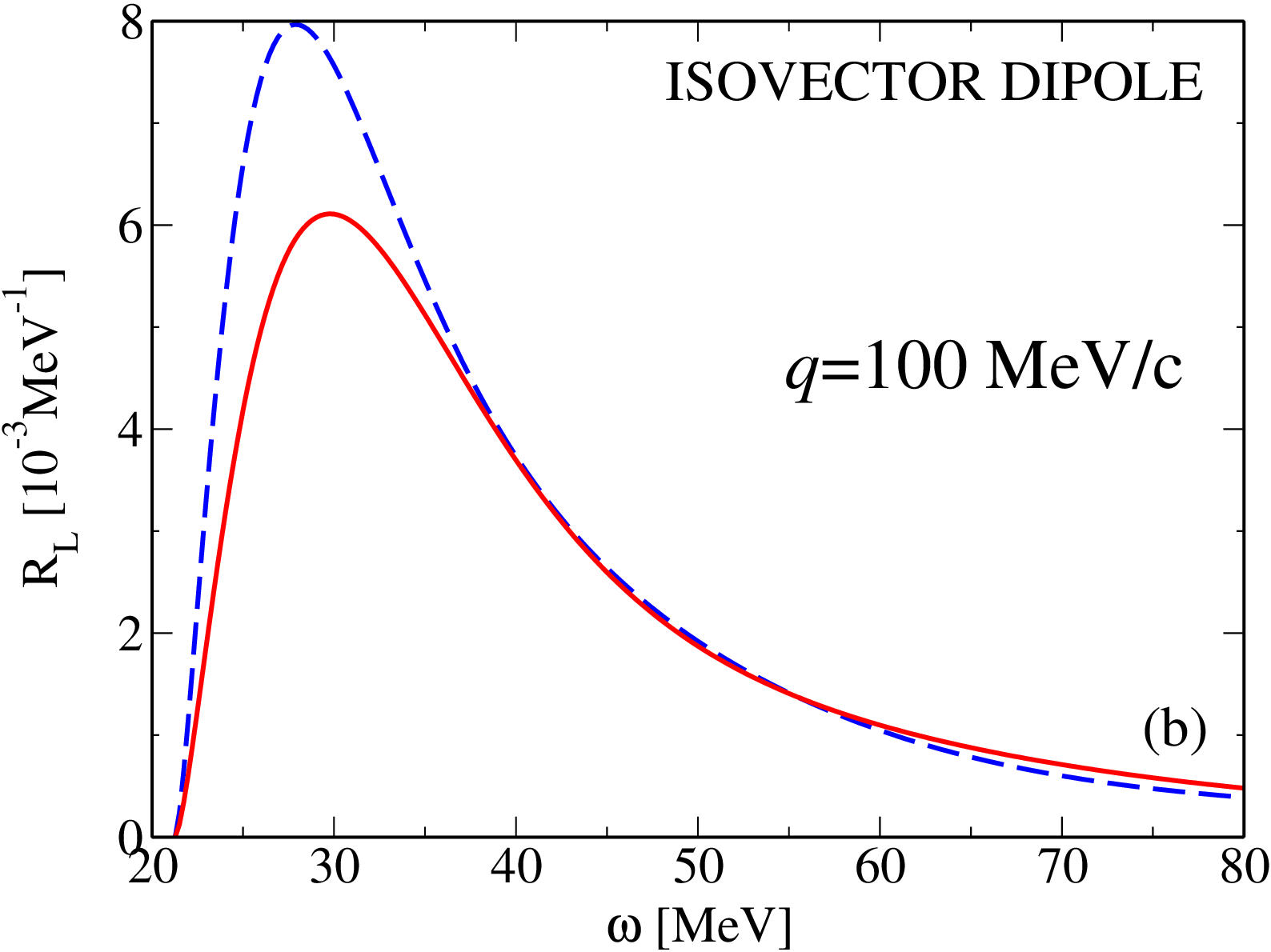}
\includegraphics[scale=0.25,clip=]{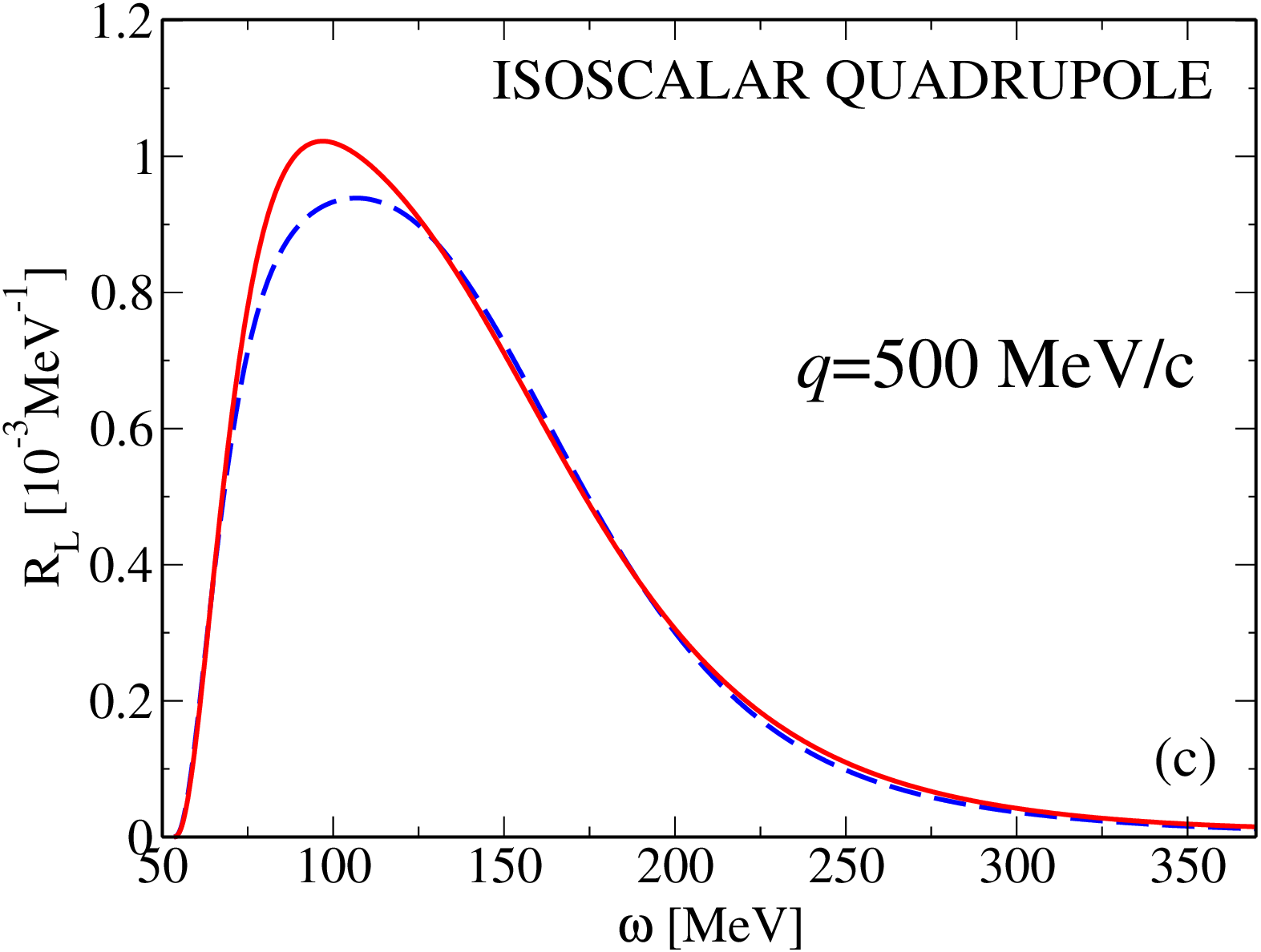}
\includegraphics[scale=0.25,clip=]{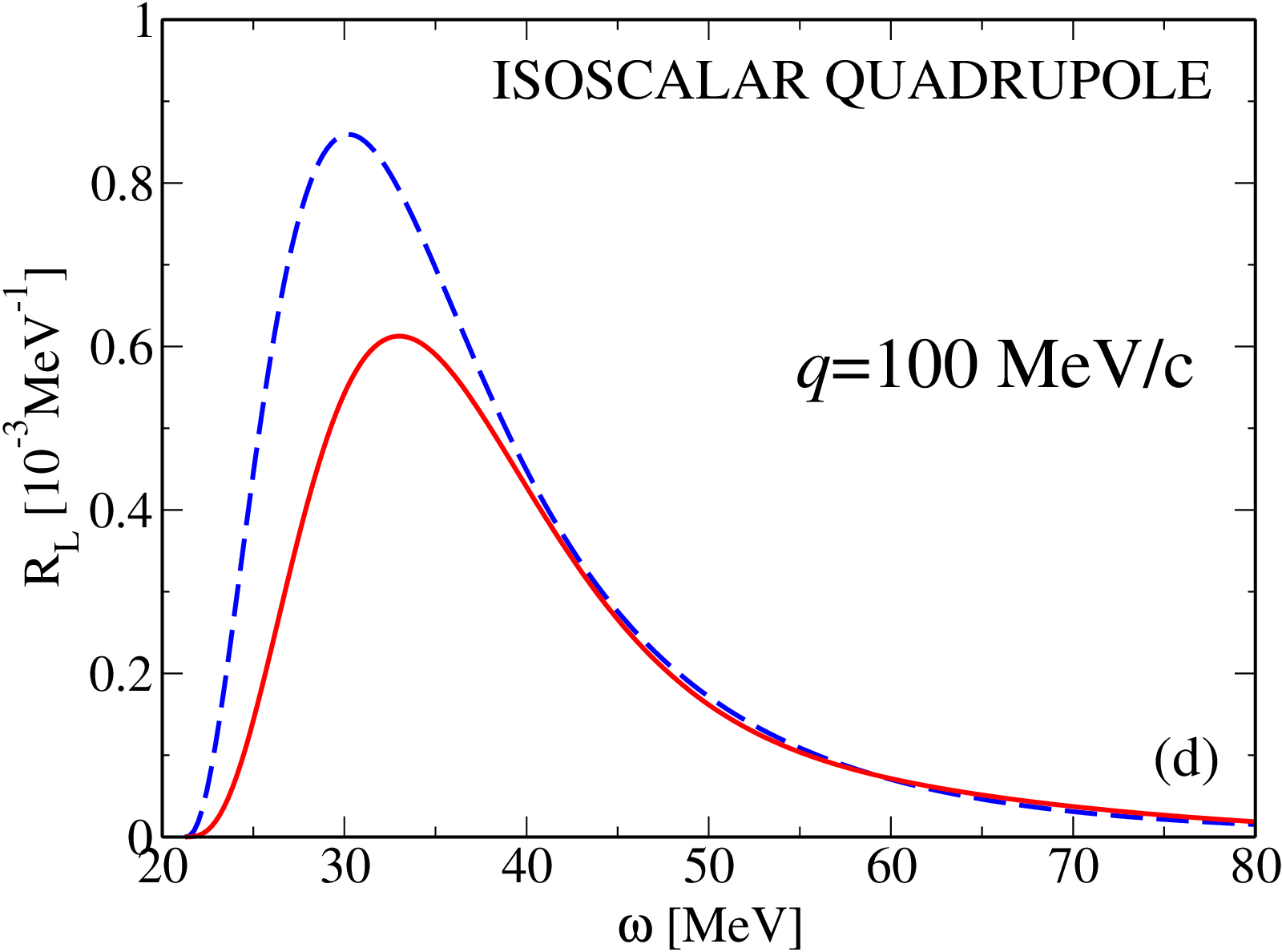}
\caption{(Color online) Response functions of the  isovector dipole and isoscalar 
quadrupole for ${q}=500$ and ${q}=100$~MeV/c with the  AV18 (dashed) and AV18+UIX 
(solid) potentials.
} 
\label{FIG6}
\end{figure*}
At $q=$100 MeV/c the $J=1$ and 3  multipoles of the isoscalar response  are tiny. 
The $J>3$ multipoles are neglected in Figs.~\ref{FIG1},~\ref{FIG2} and~\ref{FIG4} for the $q\le$100 MeV/c kinematics.
We would like to point out that the total strengths presented in Fig.~\ref{FIG5} do not contain the nucleon 
form factors. The strengths can be obtained by integrating in energy (up to infinity) the inversion of  
${\cal L}^{J,X}(\sigma,q)$ (non-energy weighted sum rule) or just by taking the norm of the right-hand-side of 
Eq.~(\ref{liteq}) for each multipole, i.e. the norm of  $\hat{C}^{J,X}(q) |{\Psi_{0}}\rangle$.

In Fig.~\ref{FIG5}, 3NF  effects are illustrated in addition.
The three shown $q$ values give an idea of 
the evolution of the effect from the short range to the long range regime.
At the highest $q$ value the three strongest contributions are given by  the isovector dipole and  the isoscalar
and isovector quadrupole. They are enhanced by the 3NF, while  all other multipoles are decreased, resulting in a net
small quenching effect. 
At $q$= 300 MeV/c one notices a kind of transition situation where only the still dominating isovector 
dipole strength is increased by the 3NF, while all other multipoles 
are quenched. At  $q= 100$ MeV/c the strength of all multipoles 
is decreased by the 3NF resulting in an overall sizable quenching effect. 

In Fig.~\ref{FIG6}, the effect described above shows up more clearly in the energy distribution of the 
dominant multipole contributions (isovector dipole, isoscalar quadrupole)
at $q=$ 500 and 100 MeV/c. In particular at $q=$ 500 MeV/c the increase
of the strength due to the 3NF in the isovector dipole  channel is found mainly in the high energy tail,
while in the 
isoscalar quadrupole the increase is found around the peak. At  $q=$ 100 MeV/c 
the situation is different in that the quenching 
of the strength due to the 3NF
concentrates in the peak region, for both multipoles.
The net result of this mechanism is the increase of the 3NF quenching  effect with decreasing $q$ that was evident in 
Fig.~\ref{FIG2}.

Here we would like to comment on the fact that the large contribution of the 3NF at low $q$ 
seems at variance with the smaller contribution to the photoabsorption cross section~\cite{Gazit06}, 
dominated as well by the isovector dipole. The reason is twofold. It has to do on the one hand with
the correct use of the Siegert theorem, and on the other hand with the common procedure to let 
theoretical cross sections start from the experimental threshold,
also when the binding energies do not reproduce the experimental values. The detailed explanation is in order here.
Because of charge conservation the relation (Siegert theorem) between the charge dipole matrix element ($C^{1,V}$) considered here  
and the electric dipole matrix element (E1) considered in the photon case implies the factor $(E_n - E_0)$ (see also \cite{ee'MT}).
The binding energy $(-E_0)$ of $^4$He, however, is about 15\% lower for AV18 than it is for AV18+UIX.
One has the following consequences. 
In the AV18+UIX case of Ref.~\cite{Gazit06} the result of the squared matrix element is multiplied by  
$(E_n - E_0)$ which is equal to $\omega_\gamma$, while in the AV18 case a multiplication by $(E_n - E_0)$  
implies a smaller multiplicative factor. Therefore the quenching three-body effect results to be smaller. 
It is only thereafter that the AV18 $^4$He total photoabsorption cross section is shifted to the experimental threshold.

One might question the procedure of shifting the theoretical cross sections to the experimental threshold, but without such a shift
all 3NF effects would be much amplified in the photoabsorption cross section  and even in the present response function results. 
However, in this case one could say that  they are in a way ``trivial binding effects''.

\section{Integral properties of $R_L$}\label{INT}

There are many examples in different fields of physics where one is not able to access a certain 
observable, but only some of its integral properties. Sum rules ($n^{th}$ moments of the energy 
distribution)~\cite{REP91} are well known examples. 
They contain a certain amount of often very useful information about the observable, but 
a limited one. The more sum rules  one knows, the larger the amount of information at disposal. Integral 
transforms can also be viewed as a special form of sum rules. While sum rules map 
an energy dependent observable into a set of $n$ discrete values, integral transforms
map the same observable into a set of continuous values.

\begin{figure}
\includegraphics[scale=0.25,clip=]{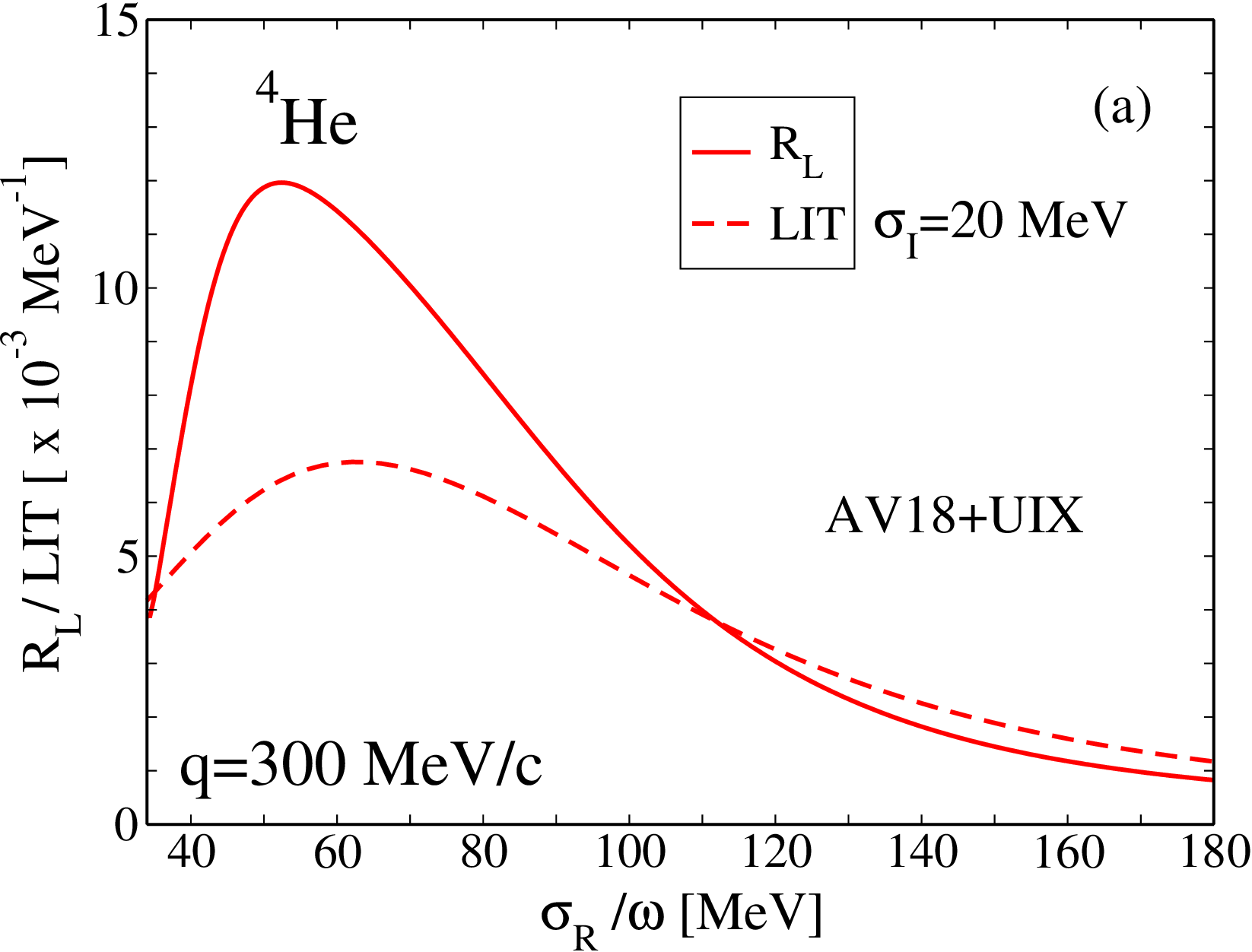}\\
~\\
\includegraphics[scale=0.25,clip=]{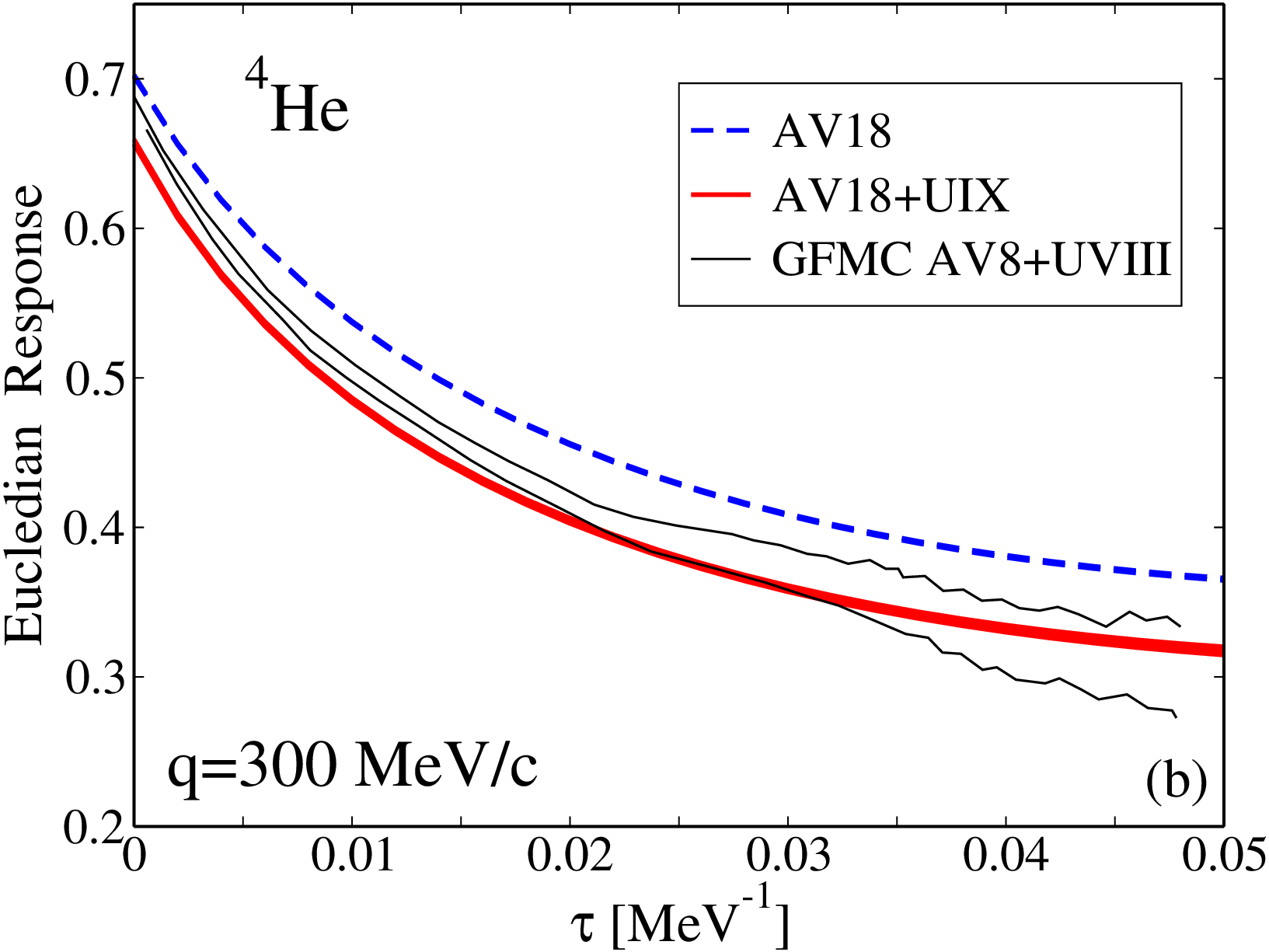}
\caption{(Color online) Longitudinal response function $R_L$ (solid) at $q=300$ MeV/c with its Lorentz Integral Transform (dashed) for the AV18+UIX potential (a). The LIT has been multiplied by $\frac{\sigma_I}{\pi}$ to normalize the integral kernel with respect to
 Eq.~(\ref{lorentz_transform}). Euclidean longitudinal
response for the same momentum transfer (b): comparison of the GFMC
calculation \cite{Eucl94} for the AV8+UVIII potential (band between thin lines) and the result of
this work with the AV18 (dashed) and AV18+UIX (solid).
}
\label{FIG7}
\end{figure}

Reconstructing the searched observable from its integral properties can be very difficult, since very 
often only a limited number of  moments are known or in case of integral transforms the 
 result of the mapping  does not resemble at all the observable of interest. This is not 
the case for the LIT. An example is illustrated in Fig.~\ref{FIG7}a 
where  we compare 
$R_L(\omega,q=300 \,{\rm MeV/c})$ for the AV18+UIX potential (as in Fig.~\ref{FIG3}e) with the corresponding LIT,
 ${\cal L}_L(\sigma,q)$, calculated using a typical value of  $\sigma_I$ (20 MeV). One notices 
the similarity between the shape of the response function and of its integral transform. This similarity is due to the fact that the 
Lorentz kernel is a representation of the $\delta$-function. It is this property that makes the inversion of the integral transform~\cite{TIK77}  reliable and 
sufficiently accurate for the Lorentz kernel.
This situation has to be confronted with the Laplace transform.
The Laplace transform of $R_L$, called Euclidean response, is given by \cite{Eucl94}
\begin{equation}
\label{lapl}
{\cal E}(\tau,q)=\int_{\omega_{th}}^{\infty}
\,d\omega\,\exp\left[-\tau \left( \omega-\frac{q^2}{2 m} \right)\right ]\frac{R_L(\omega,q)}{Z |G_E^p(Q^2)|^2}\,.
\end{equation}
Fig.~\ref{FIG7}b shows 
how $\cal E(\tau)$ exhibits a completely different form than $R_L(\omega,q)$. 
It is interesting to observe  that even in the $\tau$-space the Euclidean response obtained 
using only the two-body potential gives a result much different from that which includes the 3NF. 
However,  what is not evident from ${\cal E}(\tau)$ is in which energy region the contribution of the 3NF is important.

Fig.~\ref{FIG7}b also shows a  comparison of our results with those of Ref.~\cite{Eucl94} obtained with the Monte Carlo 
method. The comparison is of interest, even if 
the potentials used in the two cases are slightly different (in~\cite{Eucl94} the older versions of the 
Argonne and Urbana AV8 and UVIII had been used).
At $\tau$ larger than 0.02, our result with AV18+UIX
lies within the error band of the Monte Carlo numerical noise. 
At smaller $\tau$, and in particular at $\tau=0$ Fig.~\ref{FIG7}b shows a discrepancy 
between the present ${\cal E}(\tau,q)$ and that of Ref.~\cite{Eucl94}.
This is certainly due to the different potentials used.
The ${\cal E}(0)$ value corresponds to the zero-th moment of $R_L(\omega,q)$. This is 
a classical integral property of $R_L(\omega,q)$ that has been much discussed in the literature under the name of  
Coulomb sum rule (CSR) (for a review see~\cite{REP91,Ben08}). The sum rule consists in connecting the integral of the inelastic 
longitudinal response to the number of protons and to the Fourier 
transform of the proton-proton correlation function $\rho_{pp}(s)$, i.e the probability to find two 
protons at a distance $s$. In fact for the charge density operator of Eq.(~\ref{rho}) (and neglecting the neutron charge form factor)
one has

\begin{equation}
{\rm CSR}(q)\equiv\!\int_{\omega_{el}^+}\!\!\!\!\!\! d\omega
{\frac{R_L(\omega,q)}{|G_E^p(Q^2)|^2}}\! =\! Z  + Z (Z-1) f_{pp} (q) -Z^2 |F(q)|^2
\!,\label{CSR} 
\end{equation}
where $\omega_{el}$ represents the elastic peak energy,  $f_{pp} (q)$ is the Fourier 
transform of $\rho_{pp}(s)$ and $F(q)$ is the nuclear elastic form factor. (Another interesting sum rule concerning the second moment of 
$\rho_{pp}(s)$ has been considered in Ref.~\cite{tetraedro}, where one finds $\langle s^2 \rangle =  5.67$ fm$^2$  for AV18+UIX)
The main interest in CSR($q$) has been considered its  very simple model independent large $q$ limit, 
i.e. the number of protons. 
information about the proton-proton correlation function.

\begin{figure}
\includegraphics[scale=0.25,clip=]{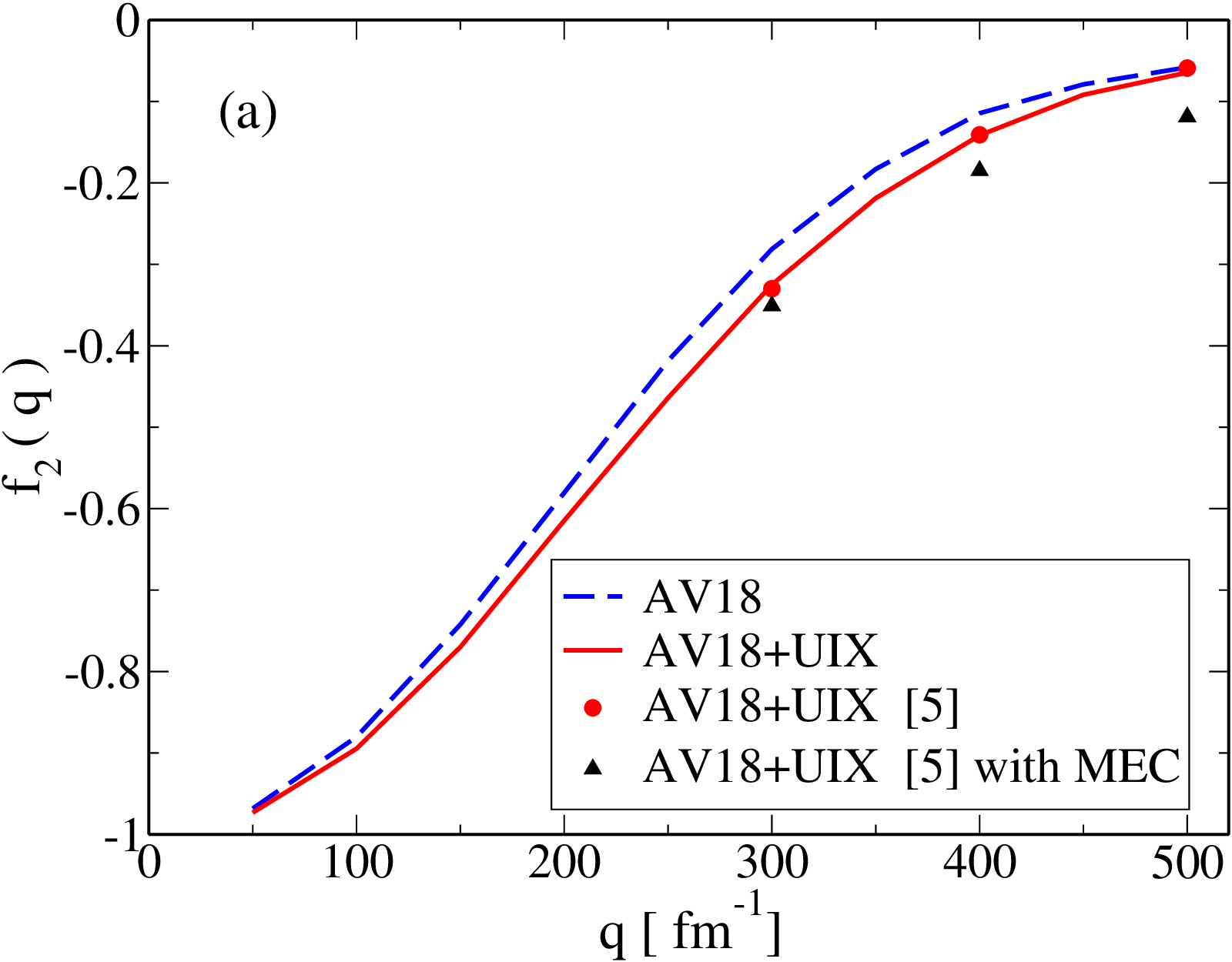}\\
~\\
\includegraphics[scale=0.25,clip=]{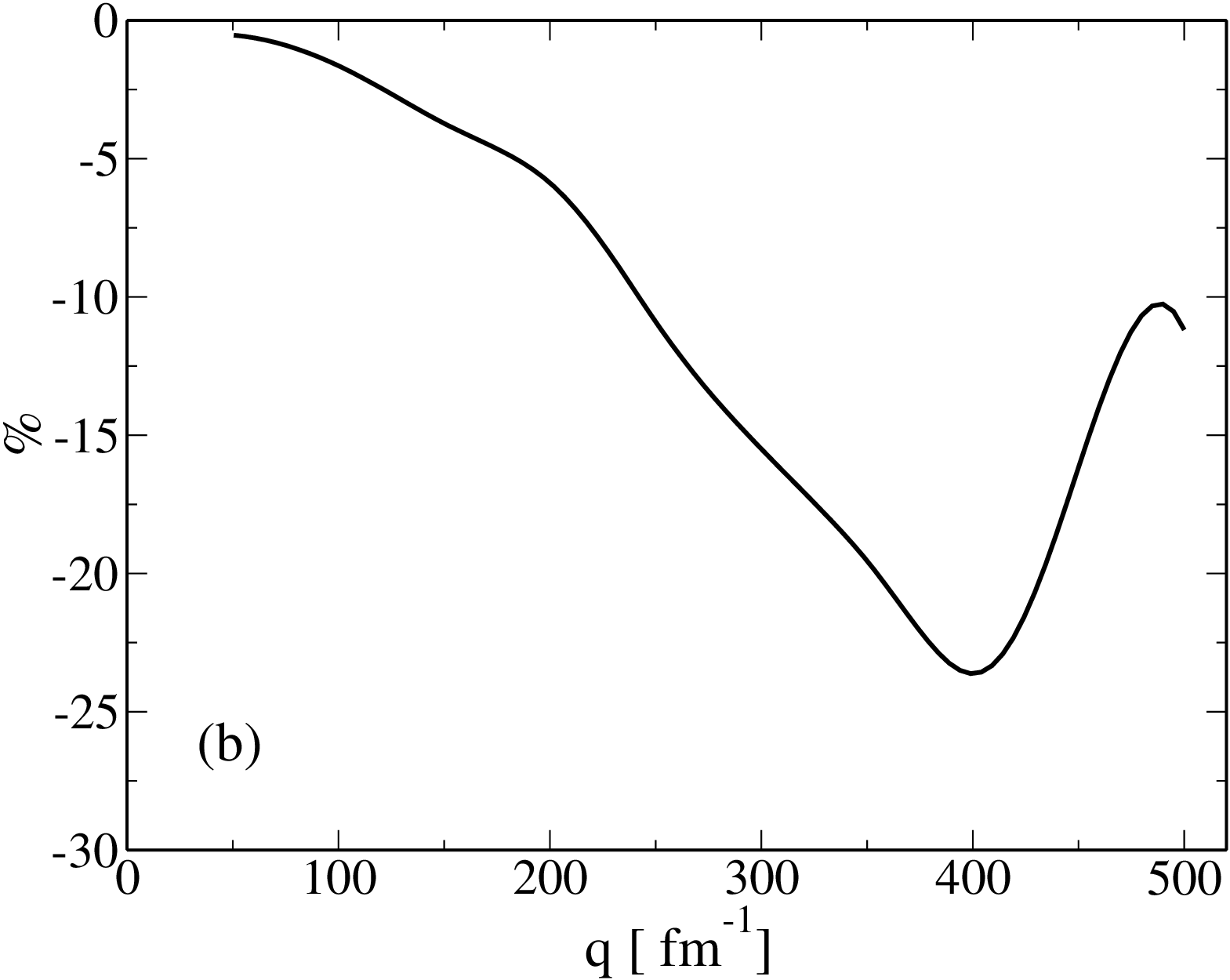}
\caption{(Color online) (a): $f_{pp}(q)$ in the CSR (Eq.~(\ref{CSR})) 
with AV18 (dashed) and AV18+UIX 
(solid) potentials.
Dots: results from~\cite{Carlson02} with the one-body density operator of Eq.(~\ref{rho}), 
triangles: results from~\cite{Carlson02} with two-body density operator.
(b): percent difference between the curves in (a) 
}
\label{FIG8}
\end{figure}

In Fig.~\ref{FIG8}, $f_{pp} (q)$ is shown in comparison with the results of Ref.~\cite{Carlson02} obtained 
with the same potential, including in addition higher relativistic order corrections as  well as exchange operators, 
neglected here. We can make the following observations: i) the perfect agreement of the two results when the operator in
Eq.~(\ref{rho}) is used shows  the great accuracy of the two calculations; ii) the contributions of   
exchange currents,  become negligible below $q= 300$ MeV. 
This means  that at low $q$ the physical interpretation of $f_{pp} (q)$ as Fourier transform of $\rho_{pp}(s)$ is safe.
Therefore, in principle, the comparison theory-experiment  would allow to study microscopically the largely unknown 
long range correlations. 
iii) as one can see in Fig.~\ref{FIG8}b the effect of the 3NF on $f_{pp} (q)$ is up to 15\% in the ``safe'' region below $q$=300 MeV/c
This gives an idea of the required experimental accuracy.
  
Unfortunately, obtaining the ``experimental`` CSR($q$) (as well as ${\cal E}(0)$) is a non trivial task, 
due to the necessity of extrapolating data up to infinite energies, even crossing the photon point, where
$(e,e')$ measurements do not have access. Different extrapolating functions have been 
proposed. They have been used also recently in Ref.~\cite{Buki06}.
Our results can help to determine these tail contributions. They can in fact be obtained subtracting 
from CSR($q$) the experimental sum of the data 
up to the last measured point at $\omega_{max}$. From the Saclay data~\cite{Saclay} at $q=$300 and 
350 MeV/c we have estimated this high energy 
contribution to be about 7\% of the CSR.
The effect becomes  twice as large for the higher $q$-values. However, 
while this procedure would be safe enough at low $q$, at large $q$ this estimate can be inaccurate 
because of the neglect of relativistic effects, two-body operators and the role of neutron form factor.
In general it would be desirable that the tail contribution does not overcome
the 3NF effect. Therefore accurate data should be taken as far in energy as possible.
Of course they cannot overcome the photon point, therefore it is interesting to calculate the contribution
of the tail beyond it. In Table~\ref{Table1}, one sees that for $q$-values up to 200 MeV/c 
the contribution of the time-like region remains very low reaching at most 1.5\%.

The previous discussion gives an idea of the experimental accuracy required to access the information about the 
proton-proton correlation function.

\begin{table}
\caption{Theoretical CSR for low $q$ with the AV18+UIX potential in
comparison with $I^{th}_{\gamma}= \int_{\omega_{el}^+}^{\omega=q} d\omega R_L^{exp}(q,\omega)/ |G_E^p(Q^2)|^2$. 
The percentage contribution of the time-like response is also shown.}
\begin{ruledtabular}
\begin{tabular}{cccc}
$q$ [MeV/c]&  CSR$(q)$  &  $I^{th}_{\gamma}$ & $\%$ time-like \\
\hline
50&   3.88 & 3.86 & 0.5 \\
100&  3.57 & 3.55 & 0.6 \\
150&  3.17 & 3.14 & 0.9 \\
200&  2.79 & 2.74 & 1.5 \\
\end{tabular}
\end{ruledtabular}
\label{Table1}
\end{table}

\section{Conclusions}\label{CONC}

In this paper we have analyzed  3NF effects on the electron scattering longitudinal response function at 
several kinematics. The most interesting results regard momentum transfers between 50 and 250 MeV/c.
Large effects of  3NFs  are found  for two different three-body potentials 
(Urbana IX and  Tucson-Melbourne$'$). We also observe that the 3NF effects differ
by non negligible amounts for the two three-body force models.
Since the difference between the two results increases with decreasing momentum transfer one can ascribe it to
rather different long range correlations generated by the the two forces. This fact
results from the difference in the continuum excitation
spectra of the two potentials. In light of this
observation one can envisage the possibility  
 to discriminate also between phenomenological and effective field theory potentials, if precise experimental data
were available at these kinematics.

Three-body force effects have been analyzed separately in the various multipoles contributing to the response.
Below $q=$ 300 MeV/c a cooperative quenching effect in all multipoles has been found.

Integral properties of the longitudinal response function have also been addressed. In particular   
the possibility to extract information about the long range behavior of the proton-proton correlation 
function has been discussed. In relation to this it has been underlined how the present results can be used in determining the 
energy tail contributions to the Coulomb sum rule.

In general it has been emphasized that different from the search for the short range correlations, the study of the long range ones
is not affected by complications due to relativistic and two-body contributions.
Therefore a Rosenbluth separation of inclusive electron scattering cross section 
of $^4$He  at momentum transfer $q\leq 200$ MeV/c would be of high value in view of a more accurate 
determination of the three-body force and in general of the long range dynamics of this system.

~\\
{\bf Acknowledgments}

We would like to thank Alexandr Buki for providing us with information about the data of Ref.\cite{Buki06}. 
This work was supported in part by the
Natural Sciences and Engineering Research Council (NSERC) and by the
National Research Council of Canada.
The work of N. Barnea was supported by the Israel Science Foundation (grant no.~361/05).
Numerical calculations were partially performed at CINECA (Bologna).


\end{document}